# Measuring and Analyzing the Shares of Economic Growth Sources in the Mining Sector of Iran: A Neoclassical Growth Accounting Approach

Dr. M. Mahmoudzadeh* S.A. Zeytoon Nejad Moosavian*
*. Department of Economics, Azad University of Iran, Firoozkooh Branch

## ABSTRACT

The purpose of this study is to measure the Total Factor Productivity (TFP) growth and determine the share of each of the economic growth sources in the mining sector of Iran. The time period of this study is 1355-1385 of the Solar Hijri calendar (roughly overlaying with the time period of 1976-2006 of the Gregorian calendar). In this paper, the shares of total factor productivity growth (TFPG) and factors' accumulations in the economic growth of the mining sector are estimated using a neoclassical growth accounting approach. Based on the estimated restricted Cobb-Douglas production function and the results obtained from the Solow residual equation, the annual growth rates of TFP were measured for each year. According to the findings, the average annual growth rate of TFP has been 2.94% during the time period of the present study. The other findings of this study indicate that the average contributions of TFPG, labor accumulation and capital accumulation in the economic growth of the mining sector have been 56%, 23%, and 21%, respectively, during the time period of the study. As such, it can be concluded that the policy of benefiting from available factors in the mining sector together with the policy of accumulating factors have simultaneously caused the value-added growth of this sector. Therefore, considering the desired performance of the mining sector in terms of its sizable productivity growth, it can be argued that the mining sector can aid Iran's economic development plans to achieve their assigned economic objectives, one of which is to increase the share of total factor productivity growth in economic growth.

# اندازه‌گیری و تحلیل منابع رشد اقتصادی بخش معدن در ایران

دکتر محمود محمودزاده* سیدعلی زیتون‌نژاد موسویان**

## چکیده

هدف این مقاله اندازه‌گیری رشد بهره‌وری کل عوامل تولید و تعیین سهم هر یک از منابع رشد اقتصادی بخش معدن، طی دورهٔ زمانی ۱۳۸۵-۱۳۵۵، می‌باشد. در این مقاله، با استفاده از روش حسابداری رشد، سهم رشد بهره‌وری و انباشت نهاده‌ها در رشد اقتصادی بخش معدن برآورد می‌شود. بر اساس مدل برآوردی تابع تولید کاب- داگلاس مقیّد و از طریق رابطهٔ باقیماندهٔ سولو، نرخ رشد سالانهٔ بهره‌وری کل عوامل برای هر سال اندازه‌گیری شد. نتایج نشان می‌دهد که میانگین نرخ رشد بهره‌وری کل عوامل طی دورهٔ بررسی، معادل ۲/۹۴٪ بوده است.

دیگر یافته‌های این مطالعه نشان می‌دهد که طی دورهٔ بررسی، رشد بهره‌وری کل عوامل، رشد نیروی کار و رشد موجودی سرمایه به‌ترتیب ۵۶٪ ، ۲۳٪ و ۲۱٪ از کل رشد اقتصادی بخش معدن را توضیح می‌دهند. با توجه به سهم بالای رشد بهره‌وری در رشد اقتصادی این بخش، می‌توان نتیجه گرفت که سیاست بهره‌گیری از منابع موجود در بخش معدن در کنار سیاست رشد عوامل تولید، موجب رشد ارزش‌افزودهٔ این بخش شده است. بنابراین، با توجه به عملکرد مطلوب بخش معدن در تحقق رشد مطلوب بهره‌وری، این بخش می‌تواند در تحقق اهداف برنامه‌های توسعه اقتصادی، مبنی بر افزایش سهم رشد بهره‌وری کل عوامل از رشد اقتصادی کشور، کمک مؤثری نماید.

## مقدمه

امروزه در اقتصادِ تمام کشورها، ارتقاء بهره‌وری به اولویتی ملّی تبدیل شده است. اقتصاددانان، ادامهٔ حیات اقتصادی کشورها را به توانایی مستمر در کسب تولید بهینه در ازای هر واحد ستاده وابسته می‌دانند و معتقدند که ارتقای بهره‌وری، رشد اقتصادی و بهبود سطح زندگی افراد جامعه را فراهم می‌آورد. بررسی مستندات برنامه‌های اول، دوم و سوم توسعهٔ اقتصادی کشور نشان می‌دهد که استراتژی توسعهٔ ملّی کشور، حصول رشد اقتصادی از افزایش در دو نهادهٔ موجودی سرمایه و اشتغال بوده است. ولی در برنامهٔ چهارم توسعه، تا حدودی دیدگاه ستاده‌گرایی جایگزین دیدگاه نهاده‌گرایی شده است. تحقق این مهم، نیازمند شناخت علمی از منابع رشد ارزش‌افزودهٔ بخش‌های اقتصادی می‌باشد که هدف این مقاله است.

بخش معدن یکی از بخش‌های مادر به شمار می‌آید که اگر چه به‌تنهایی سهم ناچیزی در تولید ناخالص داخلی دارد، اما اثر آن در تولید ثروت قابل‌ملاحظه است. بخش معدن، به عنوان عمده‌ترین منبع تأمین مواد اولیهٔ موردنیاز صنایع کشور است و از سویی، اهمیت فراوانی در ایجاد اشتغال و نیز توسعهٔ متعادل اقتصادیِ منطقه‌ای کشور دارد. اساساً بخش معدن به آن بخشی از اقتصاد اطلاق می‌گردد که اقدام به اکتشاف، استخراج و کانه‌آرایی می‌نماید. بر اساس این تعریف، اقدامات و فعالیت-هایی از قبیل ذوب، پالایش، نورد و ... مربوط به بخش صنایع معدنی بوده و در تقسیم‌بندی، جزء بخش معدن قرار نمی‌گیرند.

---

*. استادیار دانشگاه آزاد اسلامی واحد فیروزکوه Email: mahmod.ma@yahoo.com

**. مهندس معدن و کارشناس ارشد اقتصاد Email: salizey@yahoo.com

تنوع مواد معدنی در ایران به بیش از ۶۲ نوع مادهٔ معدنی بالغ می‌شود که در نوع خود کم‌نظیر است(زیتون‌نژاد، ۱۳۸۴). ایران با بیش از ۵۵ میلیارد تن ذخایر قطعی و احتمالی مواد معدنی، جزء ۱۲ کشور اول ذخیره‌دار مواد معدنی دنیا است. تنوع زیاد و فراوانی ذخایر معدنی در ایران توان بالقوهٔ زیادی برای اقتصاد کشور فراهم آورده است. اما با این همه، بخش معدن در توسعهٔ اقتصادی کشور جایگاه مناسبی ندارد(نشریه معدن و توسعه، ۱۳۸۴). عدم توجه به شاخص‌هایی چون بهره‌وری، شاید از حلقه‌های مفقوده در بهره‌برداری از فرصت‌های موجود در بخش معدن باشد.

بر اساس نظریات حسابداریِ رشد، رشد بهره‌وری کلِ اقتصاد عبارت است از مجموع رشد بهره‌وری کل در تمامی بخش-های اقتصادی. به عبارت دیگر رشد بهره‌وری کل در یک بخش می‌تواند رشد بهره‌وری کل در سطح اقتصاد، و از این طریق رشد اقتصادی را تحت تأثیر قرار دهد. تحقیقات تجربی در کشورهای توسعه‌یافته نشان می‌دهد که رشد بهره‌وری کل، بخش زیادی از رشد اقتصادی این کشورها را توضیح می‌دهد. هدف این تحقیق آن است که با اندازه‌گیری رشد بهره‌وری کل عوامل تولید در بخش معدن ایران، سهم هر یک از منابع رشد را در رشد اقتصادی این بخش اندازه‌گیری و تحلیل نماید.

سازماندهی مقاله بدین شرح است که پس از مقدمهٔ فوق، سؤالات تحقیق مطرح می‌گردد. در بخش اول، مبانی نظری بهره‌وری بررسی می‌گردد. بخش دوم به مروری بر برخی از مطالعات تجربی در زمینهٔ رشد بهره‌وری و منابع رشد، اختصاص دارد. در بخش سوم، وضعیت بخش معدن در اقتصاد ایران تحلیل می‌شود. در بخش‌های چهارم و پنجم، مدل تجربی تصریح و برآورد می‌شود. نهایتاً، بخش پایانی به جمع‌بندی و نتیجه‌گیری اختصاص دارد.
این مقاله به‌دنبال پاسخ به پرسش‌های زیر است:

- آیا رشد بهره‌وری کل عوامل تولید بر رشد اقتصادی بخش معدن تأثیرگذار بوده است؟
- سهم منابع رشد از رشد اقتصادی بخش معدن به چه ترتیب است؟

## ۱. مبانی نظری

بررسی مؤلفه‌های رشد اقتصادی در کشورهای توسعه‌یافته و درحال‌توسعه پیشرو، نشان می‌دهد که سهم «افزایش بهره-وری نیروی کار و سرمایه» غالباً چشمگیر بوده و گاه از سهم «افزایش میزان سرمایه» پیشی گرفته است(امامی میبدی، ۱۳۸۴). تجربهٔ کشورهای پیشرفته نشانگر آنست که می‌توان بخشی از رشد اقتصادی را از محل افزایش سطح بهره‌وری کل عوامل، حاصل نمود.

بهبود بهره‌وری موجب افزایش دستمزدهای حقیقی و کیفیت بهتر زندگی افراد می‌شود. از منظر تولید نیز افزایش بهره-وری، کاهش هزینه‌های تولید را به دنبال خواهد داشت. این امر نیز به نوبهٔ خود قیمت تمام شده را کاهش داده و ضمن تأمین شرایط مساعدتر برای مصرف‌کنندگان داخلی، فضای رقابت با تولیدکنندگان خارجی را نیز فراهم‌تر می‌سازد(اورعی، ۱۳۸۲). بدین ترتیب نقش بهره‌وری در بسیاری از پدیده‌ها از جمله رشد اقتصادی، ارتقاء سطح زندگی، افزایش قدرت رقابت اقتصادی، بهبود در موازنهٔ پرداخت‌های خارجی، کنترل یا مهار تورم و بسیاری از پدیده‌های دیگر، کاملاً آشکار می‌باشد.

طبق تعریف، بهره‌وری عبارت است از خارج قسمت تقسیم ستاده بر عوامل تولید. از نظر عملیاتی، بهره‌وری به معنی نسبت ستاده حقیقی به نهاده‌های حقیقی است. بر اساس این تعاریف، مفهوم بهبود یا ارتقاء سطح بهره‌وری، به معنای انتقال تابع تولید به سمت بالاتر خواهد بود(نمودار ۱).

**نمودار ۱) مقایسهٔ عملیاتی اثر رشد بهره‌وری و اثر افزایش در موجودی سرمایه**

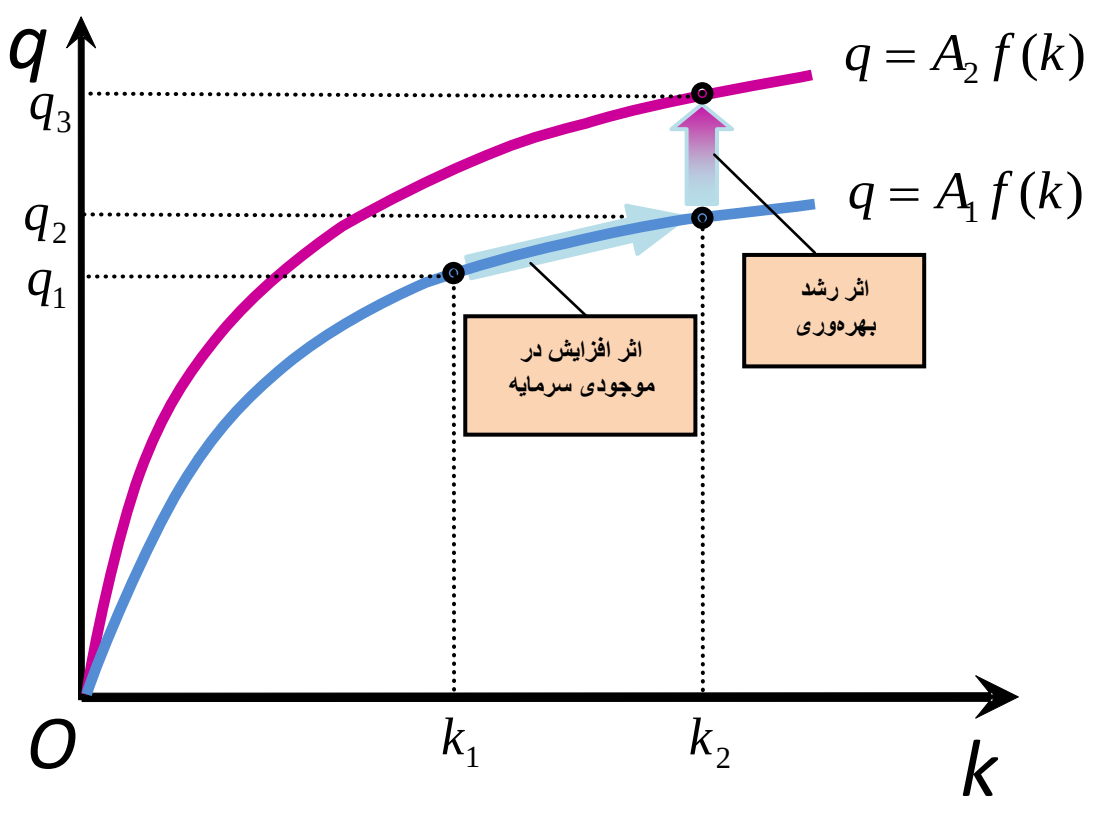

منبع: دورنبوش و فیشر(۲۰۰٤)

به‌طور کلی، شاخص‌های بهره‌وری به دو دسته شاخص‌های جزئی و کلی عوامل تولید تقسیم‌بندی می‌شوند. شاخص‌های بهره‌وری جزئی از تقسیم ارزش‌افزوده بر مقدار یک نهادهٔ معیّن به‌دست می‌آیند. اما باید توجه داشت که شاخص‌های بهره‌وری جزئی، بهره‌وری یک‌به‌یک عوامل تولید را معیّن نمی‌کنند، و از سوی دیگر، بهره‌وری کلی عوامل تولید را نشان نمی‌دهند. دیکسون[1](۱۹۹۰) به این موضوع این‌طور اشاره می‌نماید که جایگزینی سادهٔ سرمایه به‌جای نیروی کار در ترکیب نهاده‌های یک بنگاه یا یک صنعت نیز می‌تواند بهره‌وری متوسط نیروی کار را افزایش دهد. این امر بدین معناست که گاهی ممکن است افزایش بهره‌وری ناشی از جایگزینی سرمایه به جای نیروی کار(یا اصطلاحاً تعمیق سرمایه) باشد، حال آن‌که کیفیت نیروی کار اصلاً تغییری نکرده باشد. در حقیقت، این بهره‌وری کل عوامل تولید است که برآیند تغییر در کارایی تمام عوامل تولید را منظور می‌دارد. برای اندازه‌گیری بهره‌وری کل، روش‌های متفاوتی وجود دارد. یکی از این روش‌ها که معروف به روش مستقیم محاسبهٔ بهره‌وری کل عوامل تولید است، بدون استفادهٔ صریح از تابع تولید، اقدام به برآورد شاخص $TFP$ می‌نماید. در این روش، از آنجا که واحدهای اندازه‌گیری نهاده‌های بکاررفته در فرآیند تولید متفاوت هستند(نظیر نیروی کار و سرمایه)، با استفاده از تکنیک خاصی عمل جمعی‌سازی[2] نهاده‌ها را انجام داده[3]، و یک شاخص از کل نهاده‌ها ساخته می‌شود. دومین روش که مبتنی بر استفاده صریح از تابع تولید است، با در نظر گرفتن فرم تابعیِ معیّن و با انجام عملیات ریاضی بر روی تابع

1. Dixon
2. Aggregation

۳. برای جزئیات بیشتر در مورد مسئلهٔ جمعی‌سازی ر.ک. به زیتون‌نژاد(۱۳۸۷).

تولید، به برآورد شاخص $TFP$ می‌پردازد. در این روش‌ها ضمن در نظر گرفتن بعضی فروض، نرخ رشد سالانه و میانگین سالانهٔ بهره‌وری کل عوامل نیز، برآورد می‌شود.

با توجه به اینکه رشد بهره‌وری کل عوامل برابر با آن بخش از رشد تولید است که با رشد کمّی نیروی کار و سرمایه توضیح داده نمی‌شود، لذا عواملی که بهبود سطح کیفی نیروی کار و سرمایه، تخصیص بهتر منابع، استفادهٔ بهینه از منابع و امکانات موجود، را به‌همراه داشته باشد، به ارتقاء بهره‌وری کل عوامل نیز کمک می‌نماید. در ادامه و در جدول ۱، به‌طور موردی به مهم‌ترین عوامل مؤثر بر ارتقاء بهره‌وری کل عوامل اشاره می‌گردد.

**جدول ۱) عوامل مؤثر بر بهره‌وری کل عوامل**

| مهمترین عوامل | عوامل دیگر که حائز مبانی نظری نسبتاً قوی بوده و بر روی آن‌ها تحقیقات بیشتری صورت گرفته است |
|---|---|
| ۱. ارتقاء سرمایه انسانی و استفادهٔ بهینه از آن، | – تغییرات نهادی و سازمانی[۳]، |
| ۲. افزایش انگیزه نیروی کار در جهت تشویق نوآوری و خلاقیت و افزایش دقت و جدّیت افراد، | – ابداعات و ابتکارات فنی، |
| ۳. ارتقاء دانش فنی از طریق افزایش سرمایهٔ تحقیق و توسعه. | – تغییرات در سهم عوامل تولید، |
| ۴. توسعهٔ کاربرد فناوری اطلاعات و ارتباطات، | – نوسانات در تقاضا، |
| ۵. تخصیص بهتر منابع در طرح‌های سرمایه‌گذاری، | – تورم[۴]، |
| ۶. کاهش سن متوسط تجهیزات و امکانات سرمایه‌ای، | – چرخه‌های تجاری، |
| ۷. کاهش فاصلهٔ بین تولید بالفعل و تولید بالقوه، | – میزان آزاد بودن اقتصاد[۵]، |
| ۸. استفادهٔ بهینه از انرژی و نهاده‌های واسطه‌ای و بهبود در کیفیت نهاده‌های واسطه‌ای، | – سرمایه‌گذاری مستقیم خارجی[۶] (*FDI*)[۷]، |
| ۹. بکارگیری روش‌های نوین مدیریتی، | – مقیاس تولید، |
| ۱۰. کاهش نرخ تورم به‌منظور تخصیص بهینهٔ منابع بین بخش‌های اقتصادی[۱]، | – زیرساخت‌ها[۸]، |
| ۱۱. گسترش شرایط رقابت‌پذیری اقتصاد از راه‌هایی نظیر افزایش درجهٔ باز بودن اقتصاد[۲] و خصوصی‌سازی. | – ساختار سنی[۹]، |
| | – تعرفه‌ها، |
| | – و... . |

منبع: جمع‌بندی نویسندگان.

[۱]. برای اطلاع بیشتر ر.ک. به خالصی(دی۱۳۸۳)؛ Bitors & Panas(2001)

[۲]. برای اطلاع بیشتر ر.ک. به خالصی(دی۱۳۸۳)؛ Miller & Upadhyay(2000)

[۳]. برای جزئیات بیشتر ر.ک. به Bennedsen M. et al. (2005)

[۴]. تورم به‌واسطهٔ به‌وجود آوردن عدم اطمینان به آینده، از بین بردن اعتماد نسبت به اقتصاد و بی‌ثباتی در جامعه موجب کاهش بهره‌وری شده و اساساً همبستگی منفی(رابطهٔ معکوس) با آن دارد. برای جزئیات بیشتر ر.ک. به Min L. (2006)

[۵]. برای نمونه ر.ک. به Bessonova, E. et al. (2002)

[۶]. برای نمونه ر.ک. به Khan, S. (2006)

[7]. Foreign Direct Investment

[۸]. برای نمونه ر.ک. به Wylie, P. (1997) و Ashok, K. and Balasubramanian R. (2006)

[۹]. برای نمونه ر.ک. به Kogel, T. (2003)

عموماً متخصصان زمینه‌های مختلف از روش‌های متفاوتی جهت اندازه‌گیری بهره‌وری استفاده می‌نمایند. در این میان اقتصاددانان از روش‌های عددشاخص[1]، تابع مسافت[2] ، اقتصادسنجی[3] و حسابداری رشد[4] استفاده می‌نمایند. در رویکرد عدد شاخص جهت محاسبهٔ بهره‌وری، یک شاخص مقدار ستاده به یک شاخص مقدار نهاده تقسیم می‌گردد، تا اینکه یک شاخص بهره‌وری ارائه شود.

رویکرد تابع مسافت اساساً درصدد آن است که رشد $TFP$ را به دو جزء مختلف تفکیک نماید. در اصل، این تکنیک این امکان را فراهم می‌سازد که هرگونه تغییر در $TFP$ را به تغییر ناشی از حرکت به سمت مرز تولید و انتقال مرز، تفکیک نمود.

منطق رویکرد اقتصادسنجی در اندازه‌گیری بهره‌وری، برآورد نمودن پارامترهای تابع تولید(یا تابع هزینه یا تابع سود) می‌باشد. رویکرد اقتصادسنجی در اندازه‌گیری بهره‌وری صرفاً بر اساس مشاهدات حجم ستاده‌ها و نهاده‌ها قرار گرفته است. در این رویکرد از برقرار نمودن قیاس منطقی و برقراری ارتباط بین کشش‌های تولید و سهم‌های درآمدی(که ممکن است مطابق با واقعیت باشند و یا نباشند) اجتناب می‌شود. این امر در حقیقت محققین را در موضع آزمون‌نمودن این رابطه‌ها و ارتباطات قرار می‌دهد(سازمان همکاری‌های اقتصادی و توسعه، ۲۰۰۱).

یکی از مزایای برجستهٔ رویکرد اقتصادسنجی در اندازه‌گیری بهره‌وری، وجود امکان اطلاع یافتن از وضعیت کامل تکنولوژی تصریح‌شده می‌باشد. در این رویکرد، علاوه بر برآوردهای بهره‌وری، اطلاعاتی در مورد دیگر پارامترهای تکنولوژی تولید نیز به‌دست می‌آید. به‌علاوه، به دلیل اینکه رویکرد اقتصادسنجی بر پایهٔ استفاده از اطلاعات ستاده‌ها و نهاده‌ها استوار است، لذا این رویکرد در تصریح تکنولوژی تولید، دارای انعطاف‌پذیری بیشتری می‌باشد. به‌عنوان مثال، در این رویکرد، امکان معرفی دیگر فرم‌های تغییرات تکنولوژیکی عامل- اندوز[5] (فرم‌هایی بجز فرم خنثی هیکس که در رویکردهای عدد شاخص و حسابداری رشد، تصریح می‌شوند) را مقدور می‌سازد(ماوسون و همکاران[6]، ۲۰۰۳). در چارچوب مفهومی اقتصادسنجی، امکان آزمون نمودن اعتبار فروضی که اساس رویکردهای عدد شاخص و حسابداری رشد را تشکیل می‌دهند نیز وجود دارد. به عنوان مثال، وجود قابلیت آزمون‌نمودن فرض بازدهی ثابت نسبت به مقیاس، که اغلب در رویکرد حسابداری رشد مورد استفاده قرار می‌گیرد، اشاره نمود. هولتن[7](۲۰۰۰) با استناد به چند دلیل، توضیح می‌دهد که رویکرد اقتصادسنجی در اندازه‌گیری بهره‌وری وری باید به عنوان مکملی برای رویکردهای عدد شاخص و حسابداری رشد، در نظر گرفته شود. یکی از آن دلایل این است که، سادگی نسبی رویکردهای عدد شاخص و حسابداری رشد می‌تواند به تفسیر بهتر نتایج پرمایهٔ حاصل از رویکرد

1. Index Approach
2. Distance Function Approach
3. Econometrics Approach
4. Growth Accounting Approach
5. Factor-augmenting
6. Mawson et al.
7. Hulten

اقتصادسنجی کمک نماید(اسکریر و پیلات[1]، ۲۰۰۱). در نهایت می‌توان اظهار داشت که رویکرد اقتصادسنجی مناسب‌ترین ابزار در مطالعات انفرادی[2] و مطالعات دانشگاهی‌محور[3](علمی‌محور) می‌باشد(سازمان همکاری‌های اقتصادی و توسعه، ۲۰۰۱).

حسابداری رشد رویکردی است که رشد اقتصادی مشاهده‌شده را به دو جزء مجزای تغییرات در نهاده‌های عوامل و جزءِ پسماند تفکیک می‌نماید. اساساً جزءِ پسماند[4] بیانگر تغییرات در پیشرفت فنّی و سایر عوامل می‌باشد. به‌طور کلی، انجام حسابداری رشد، به‌عنوان گام اولیه در تحلیل عوامل اصلی تعیین‌کنندۀ رشد اقتصادی مدنظر قرار می‌گیرد. متعاقباً، گام نهایی شامل برقراری رابطه‌ای بین نرخ‌های رشد موجودی عوامل، سهم‌های عوامل و تغییر تکنولوژیکی(همان عامل پسماند) با عناصری از قبیل سیاست‌های دولت، ترجیحات خانوار، منابع طبیعی، سطوح اولیۀ سرمایۀ فیزیکی و انسانی و عواملی از این دست خواهد بود. اجرای حسابداری رشد، خصوصاً هنگامی می‌تواند مفید باشد که عوامل اصلی تعیین‌کننده‌ای که در نرخ‌های رشد عوامل اهمیت دارند، اساساً مستقل از آنهایی باشند که در تغییر تکنولوژیکی اهمیت دارند(بَرو[5]، ۱۹۹۸).

مبانی اولیۀ حسابداری رشد در مقالات سولو[6](۱۹۵۷)، کندریک[7](۱۹۶۱)، دنیسون[8] (۱۹۶۲) و جورگنسن و گریلیچس[9](۱۹۶۷) ارائه گردید. گریلیچس در بخش اول مقالۀ سال ۱۹۹۷ خود، با تأکید بر لزوم توسعۀ باقیماندۀ سولو، مروری بر این تاریخچۀ فکری ارائه می‌نماید.[10] حسابداری رشد رویکردی است که از طریق آن، امکان تفکیک مکانیکی رشد تولید(به دو دستۀ کلی رشد نهاده‌های مختلف و تغییرات سطح بهره‌وری کل عوامل تولید)، فراهم می‌گردد. استفاده از رویکرد حسابداری رشد مستلزم تصریح تابع تولید می‌باشد.[11] این تابع بیانگر آن است که در زمان‌های مختلف، با فرض در دست بودن سطح مشخصی از نهاده‌ها و بهره‌وری کل عوامل، چه سطحی از محصول را می‌توان تولید نمود. بر این اساس تابع تولید نئوکلاسیکی به صورت زیر نوشته می‌شود:

$$Y = f(A, K, L) \quad [۱]$$

که در آن $Y$ بیانگر میزان تولید، $A$ سطح تکنولوژی، $K$ موجودی سرمایه و $L$ نیروی کار می‌باشد. بدیهی است که موجودی سرمایه و نیروی کار هر کدام به‌تنهایی می‌تواند به انواع و یا کیفیت‌های مختلفی تقسیم شوند. به عنوان نمونه‌ای از این

---

1. Schreyer and Pilat
2. Single studies
3. Academically oriented
4. Residual
5. Barro
6. Solow
7. Kendrick
8. Denison
9. Jorgenson and Griliches

۱۰. البته سولو، خود نیز در جمله‌ای معروف بر اهمیت این موضوع صحه گذاشته است. وی می‌گوید: "...و امّا بخشی از رسالت علم اقتصاد، غربال نمودن جملات خطاست. با اینکه این امر دشوارتر از به‌دست آوردن آن‌ها می‌باشد، ولیکن مطمئناً اقدام جالب‌تری خواهد بود."

۱۱. زیتون نژاد (a۲۰۱۶، b۲۰۱۶، و ۲۰۱۷) نقش توابع تولید به عنوان بخش مهمی از بنیان های خرد اقتصاد کلان را توضیح می دهد. نامنکو و زیتون نژاد (۲۰۱۶ و ۲۰۱۷) تئوری تولید و نقش تابع تولید در آن را به صورت گرافیکی و جبری توضیح می دهند.

اقدامات می‌توان به جورگنسن و گریلیچس(۱۹۶۷) اشاره نمود. رویکرد حسابداری رشد بر پایهٔ چندین فرض مهم استوار است. این فروض عبارتند از:

**۱.** فرض اول این است که جملهٔ تکنولوژی یا همان جملهٔ بهره‌وری کل عوامل($A$)، قابل‌تفکیک باشد. به‌عبارت دیگر، باید بتوان معادلهٔ [۱] را به صورت $Y=A.f(K,L)$ نوشت.

**۲.** فرض می‌گردد تابع تولید دارای خصوصیت بازدهی ثابت نسبت به مقیاس باشد.

**۳.** فرض می‌شود که تولیدکنندگان به‌طور کارا رفتار می‌کنند. این فرض بدین معناست که آنها در تلاش برای حداکثرسازی سود می‌باشند.

**۴.** فرض می‌شود که بازارها به‌طور کامل رقابتی بوده و تمام تولیدکنندگان قیمت‌پذیر می‌باشند. در نتیجه، تنها مقدارها را تعدیل نموده و هیچ‌گونه اثر انفرادی بر قیمت‌ها ندارند.

در ادامه به معرفی معادلات مطرح در این رویکرد، و نحوهٔ استخراج آنها اشاره خواهد شد. در صورتی که قبل از اعمال فرض تکنولوژی خنثی هیکس، از رابطهٔ اولیهٔ $Y=f(A,K,L)$ نسبت به زمان مشتق گرفته و معادلهٔ حاصله را بر $Y$ تقسیم نموده و به‌منظور نیل به اهداف خاصی، $K$ و $L$ را در صورت و مخرج دو جزءِ سمت راست معادلهٔ نهایی، ضرب نماییم خواهیم داشت:

$$\frac{\dot{Y}}{Y}=g+(\frac{F_K\cdot K}{Y})\cdot(\frac{\dot{K}}{K})+(\frac{F_L\cdot L}{Y})\cdot(\frac{\dot{L}}{L}) \qquad [۲]$$

که در آن $F_K$ و $F_L$ تولید نهایی (اجتماعی) عوامل تولید و $g$ مقدار رشد ناشی از تغییر تکنولوژی بوده و در حقیقت به‌صورت زیر مطرح می‌شود:

$$g\equiv(\frac{F_A\cdot A}{Y})\cdot(\frac{\dot{A}}{A}) \qquad [۳]$$

در این شرایط اگر عامل تکنولوژی برحسب حالت تکنولوژی خنثی هیکس ظاهر شود، خواهیم داشت $F(A,K,L)=A\cdot\tilde{F}(K,L)$، در نتیجه $g=\frac{\dot{A}}{A}$ خواهد شد. بنابراین می‌توان نرخ پیشرفت تکنولوژیکی را با استفاده از معادلهٔ قبل، به عنوان یک پسماند محاسبه نمود:

$$g=\frac{\dot{Y}}{Y}-(\frac{F_K\cdot K}{Y})\cdot(\frac{\dot{K}}{K})-(\frac{F_L\cdot L}{Y})\cdot(\frac{\dot{L}}{L}) \qquad [۴]$$

بسیار واضح است که معادلهٔ بالا قابل‌سنجش نمی‌باشد، چرا که برآورد این معادله مستلزم اطلاع داشتن از مقادیر تولید نهایی اجتماعی( $F_K$ و $F_L$ ) است. بر این اساس، در عمل، محاسبات نوعاً فرض می‌کنند که می‌توان تولید نهایی جامعه را از طریق مشاهدهٔ قیمت‌های عوامل اندازه‌گیری نمود. در صورتی که عوامل تولید، معادل تولید نهایی خودشان را دریافت کنند،

به‌طوری که $F_K = R$ (قیمت اجارۀ سرمایه) و $F_L = w$ (نرخ دستمزد) باشد، بنابراین برآورد اولیۀ استاندارد مربوط به پیشرفت تکنولوژیکی، بر اساس معادلۀ قبل، به‌صورت معادلۀ زیر خواهد بود:

$$\hat{g} = \frac{\dot{Y}}{Y} - s_K \cdot (\frac{\dot{K}}{K}) - s_L \cdot (\frac{\dot{L}}{L}) \quad [۵]$$

که در آن $s_K \equiv \frac{RK}{Y}$ و $s_L \equiv \frac{wL}{Y}$ سهم‌های پرداخت به هر عامل تولید از کل ستاده می‌باشد. از مقدار عددی $\hat{g}$ اغلب به‌عنوان برآوردی از رشد بهره‌وری کل عوامل یا باقیماندۀ سولو[1]، یاد می‌شود(بَرو، ۱۹۹۸). جهت استفاده از معادلۀ اخیر و برای به‌دست آوردن رشد بهره‌وری کل عوامل، صرفاً کافیست که برآوردهای $s_K$ و $s_L$ را در دست داشته باشیم. اساساً روش‌های مختلفی برای محاسبۀ $s_K$ و $s_L$ وجود دارد که در ادامۀ مباحث این بخش به آن‌ها پرداخته خواهد شد.

نکتۀ مهم در برآوردهای رشد $TFP$ به‌دست‌آمده از طریق بکارگیری مستقیم رابطۀ $\hat{g} = \frac{\dot{Y}}{Y} - s_K \cdot (\frac{\dot{K}}{K}) - s_L \cdot (\frac{\dot{L}}{L})$، این است که در این برآوردهای رشد $TFP$ ، از برآوردهای اقتصادسنجی استفاده نمی‌شود. در این برآوردها، پسماند برآوردشده( $\hat{g}$ ) در هر مقطع زمانی، صرفاً با استفاده از داده‌های سری‌زمانی $\frac{\dot{Y}}{Y}$، $\frac{\dot{K}}{K}$، $\frac{\dot{L}}{L}$، $s_K$ و $s_L$ محاسبه می‌گردد. محققان، در عمل، میانگین مقادیر عددی $\hat{g}$های محاسبه‌شده را برای یک دورۀ زمانی معیّن، گزارش می‌کنند.

در برآورد سهم عوامل تولید از روش‌های رگرسیون و حسابداری ملّی استفاده می‌شود. در روش رگرسیون، سهم عوامل با نسبت دادن نرخ رشد ستانده به نرخ رشد هر یک از نهاده‌ها و یک مقدار ثابت، تخمین زده می‌شود. روش دوم، داده‌های حاصل از آمارهای حساب‌های ملّی را برای تخمین سهم‌های عوامل، از طریق اندازه‌گیری سهم درآمدی هر عامل تولید، مورد استفاده قرار می‌دهد. صرف نظر از نوع روش به‌دست آوردن سهم عوامل، سهم منابع رشد از رشد اقتصادی به شرح زیر به‌دست می‌آید:

| |
|---|
| **ضریب عامل سرمایه در تابع رشد تولید × متوسط رشد سالیانه عامل سرمایه = سهم رشد عامل سرمایه** |
| **ضریب عامل نیروی‌کار در تابع رشد تولید × متوسط رشد سالیانه نیروی‌کار = سهم رشد عامل نیروی‌کار** |
| **سهم‌عامل‌رشدسرمایه + سهم‌رشدعامل‌نیروی‌کار) – متوسط رشد سالیانه ستاده = سهم رشد *TFP*** |

بسیاری از محققین، حسابداری رشد را به‌عنوان ابزاری ضروری در مدیریت رشد و دارای رهاوردهایی در زمینۀ مباحث تاریخ اقتصادی، همچون بررسی سیر تحولات سهم عوامل تعیین‌کنندۀ رشد از رشد اقتصادی، می‌دانند. برای نمونه می‌توان به کِرَفتس[2](۲۰۰۸) اشاره نمود.

[1]. Solow Residual
[2]. Crafts

## ۲. ادبیات تجربی

علیمرادی(۱۳۸۲) در تحقیقی به اندازه‌گیری رشد بهره‌وری کل عوامل تولید در سطح اقتصاد کشور با استفاده از اطلاعات دوره زمانی ۱۳۷۹-۱۳۴۵ اقدام نموده است. وی تابع تولید اقتصاد کشور را با در نظر گرفتن توابع تولید کاب- داگلاس[۱] و ترنسلوگ[۲] برآورد نموده و به این نتیجه رسیده است که متوسط سهم رشد سالیانهٔ بهره‌وری کل در رشد اقتصادی کشور طی دوره مورد بررسی برابر۸/۱۲- درصد می‌باشد. این امر بیانگر تأثیر منفی بهره‌وری کل بر رشد اقتصادی کشور در طی دورهٔ فوق می باشد.

شاه‌آبادی(۱۳۸۵) در مطالعه‌ای دیگر به بررسی منابع رشد بخش صنایع و معادن اقتصاد ایران با استفاده از اطلاعات سری زمانی ۱۳۵۶-۱۳۴۲(قبل از انقلاب) و ۸۳-۱۳۶۸(بعد از انقلاب) پرداخته است. وی با استفاده از آزمون همگرایی[۳] به بررسی وجود رابطهٔ بلندمدت معنادار بین متغیرها پرداخته و در نهایت با استفاده از روش یوهانسن[۴] ضرایب معادلهٔ تولید بخش صنایع و معادن را برآورد نموده است. وی معتقد است منابع رشد بخش صنایع و معادن قبل از انقلاب به‌ترتیب موجودی سرمایهٔ فیزیکی، نیروی کار و بهره‌وری کل عوامل بوده و بعد از انقلاب به‌ترتیب موجودی سرمایهٔ فیزیکی، بهره‌وری کل عوامل و نیروی کار می‌باشد.

سبحانی و عزیزلو(۱۳۸۷) به تحلیل مقایسه‌ای بهره‌وری عوامل تولید در زیربخش‌های صنایع بزرگ ایران پرداخته‌اند. مطالعهٔ بهره‌وری کلی عوامل تولید در قالب تخمین توابع تولید زیربخش‌های صنایع بزرگ، حاکی از این امر است که بیشترین نرخ رشد بهره‌وری عوامل تولید به‌ترتیب به زیربخش‌های صنایع ماشین‌آلات، تجهیزات، ابزار و محصولات فلزی، صنایع تولید فلزات اساسی، صنایع شیمیایی، صنایع کاغذ، مقوا، چاپ و صحافی و صنایع چوبی و محصولات چوبی، اختصاص دارد.

لیانگ[5](۲۰۰۱) در تحقیقی به اندازه‌گیری بهره‌وری کل عوامل در جمهوری خلق چین پرداخته به این نتیجه رسیده است که منابع رشد اقتصادی این کشور طی سال‌های ۱۹۹۶-۱۹۶۱، به‌ترتیب عبارت از رشد موجودی سرمایه، رشد موجودی نیروی کار و افزایش بهره‌وری کل عوامل بوده است. وی در ادامه نتیجه می‌گیرد که با گذر زمان سهم رشد *TFP* در رشد اقتصادی افزایش یافته است.

کلاسک[6] و همکاران (۲۰۰۸) با استفاده از تابع تولید ترنسلوگ چهار نهاده‌ای به اندازه‌گیری سطح و رشد بهره‌وری کل عوامل در صنعت تولیدات کارخانه‌ای کشور چک پرداخته‌اند. بر اساس نتایج به‌دست‌آمده، در دورهٔ ۲۰۰۶-۱۹۹۵، سهم رشد *TFP* از رشد ستاده در بخش تولیدات کارخانه‌ای چک ۱۳درصد بوده است. این سهم در نیمهٔ دوم دههٔ ۱۹۹۰، ۱۱درصد بوده است و در دورهٔ ۲۰۰۶-۲۰۰۳ به ۱۶/۲درصد افزایش پیدا کرده است.

---

۱. Cobb-Douglas
۲. Translog
۳. Cointegration Test
۴. Johansen
5. Liang
6. Klacek

جورگنسن و استیرو[1](۲۰۰۰) به بررسی بهره‌وری ایالات متحده برحسب صنعت پرداخته‌اند. روش اصلی این تحقیق حسابداری رشد بوده و در این راستا برای هر صنعت از یک تابع تولید کل منحصربه‌فرد استفاده شده است. نتایج این تحقیق نشان می‌دهد که بهره‌وری سطح اقتصاد ایالات متحده در طول دورهٔ زمانی ۱۹۹۶-۱۹۵۸ به‌طور متوسط، سالانه به میزان ۰/۴۵ درصد افزایش یافته است.

در ادامه به‌منظور جمع‌بندی اهم مطالعات انجام‌شده در زمینهٔ اندازه‌گیری رشد بهره‌وری کل عوامل و بررسی منابع رشد اقتصادی، خلاصهٔ مطالعات تجربی صورت‌گرفته به‌صورت یک جدول ارائه می‌گردد.

**جدول ۲) خلاصه مطالعات تجربی در زمینهٔ اندازه‌گیری تغییرات بهره‌وری کل عوامل و بررسی تأثیر آن بر میزان ستاده**

| نویسنده | کشور | دورهٔ مطالعه | سطح مطالعه | نتیجه* |
|---|---|---|---|---|
| علیمرادی (۱۳۸۲) | ایران | ۱۳۷۹-۱۳٤٥ | کلان | - |
| شاه آبادی (۱۳۸٥) | ایران | ۱۳٥٦-۱۳٤۲ و ۱۳۸۳-۱۳٦۸ | بخش صنایع و معادن ایران | + |
| کمیجانی و صلاحی (۱۳۸٦) | ایران | ۱۳۸٥-۱۳۷۳ | زیربخش صنایع معدنی | - |
| عباسیان و مهرگان (۱۳۸٦) | ایران | ۱۳۷۹-۱۳٤٥ | بخش‌های اقتصادی | + - |
| سبحانی و عزیزلو (۱۳۸۷) | ایران | ۱۳۸۳-۱۳٥۰ | زیربخش‌های صنایع بزرگ | + - |
| قلی‌زاده و صالح (۱۳۸٤) | ایران | ۱۳۸۱-۱۳٥۷ | کشاورزی | + |
| اکبری و رنجکش (۱۳۸۲) | ایران | ۱۳۷٥-۱۳٤٥ | کشاورزی | + |
| امینی (۱۳۸٦) | ایران | ۱۳۸۲-۱۳۷۰ | بخش‌های اقتصادی | + - |
| محمودزاده، بابازاده و زیتون‌نژاد (۱۳۸۸) | ایران | ۱۳۸٥-۱۳٥٥ | بخش معدن | + |
| سولو (۱۹٥۷) | آمریکا | ۱۹٤۹-۱۹۰۹ | کلان | + |
| دنیسون (۱۹۷٤) | آمریکا | ۱۹٦۹-۱۹۲۹ | کلان | + |
| لیانگ (۲۰۰۱) | چین | ۱۹۹٦-۱۹٦۱ | کلان | + |
| جورگنسن و گالوپ (۱۹۷٦) | آمریکا | ۱۹۷۳-۱۹٤۷ | کلان بر حسب صنایع | + - |
| کلاسک و همکاران (۲۰۰۸) | چک | ۲۰۰٦-۱۹۹٥ فصلی | بخش تولیدات کارخانه‌ای | + |
| جورگنسن و استیرو (۲۰۰۰) | آمریکا | ۱۹۹٦-۱۹٥۸ | کلان بر حسب صنایع | + - |

*. + بیانگر رشد بهره‌وری کل عوامل، - بیانگر رشد منفی بهره‌وری کل عوامل و +- بیانگر تفاوت وضعیت رشد در بخش‌ها یا صنایع مختلف می‌باشد. منبع: جمع‌بندی نویسندگان.

## ۳. وضعیت بخش معدن در اقتصاد ایران

ایران با در اختیار داشتن حدود ۵۵ میلیارد تن ذخایر معدنی، براساس آمار سال ۲۰۰۴ و برحسب معیارهای وزنی، ۱/۲۴٪ از تولیدات مواد معدنی دنیا را انجام می‌دهد(زیتون‌نژاد ۱۳۸۴). نتایج آمارگیری سال ۱۳۸۶ معادن کشور نشان می‌دهد که در سال ۱۳۸۵ تعداد ۳۵۸۲ معدن درحال بهره‌برداری در کشور وجود داشته است و مجموعاً ۶۰۰۶۲ نفر در معادن درحال بهره‌برداری کشور به‌کار اشتغال داشته‌اند(مرکز آمار ایران، ۱۳۸۷). داده‌های بانک مرکزی نشان می‌دهد که طی دورهٔ زمانی ۱۳۸۵-۱۳۵۵، متوسط سهم نسبی بخش معدن از تولید ناخالص داخلی، بسیار اندک و در حدود نیم درصد بوده است. این در حالیست که این سهم در بسیاری از کشورهای پیشرفته به‌مراتب بیش از این میزان می‌باشد(نمودار ۲).

[1]. Jorgenson and Stiroh

نمودار ۲) سهم بخش معدن در تولید ناخالص ملی کشورهای مختلف بر اساس داده‌های سال ۲۰۰۴

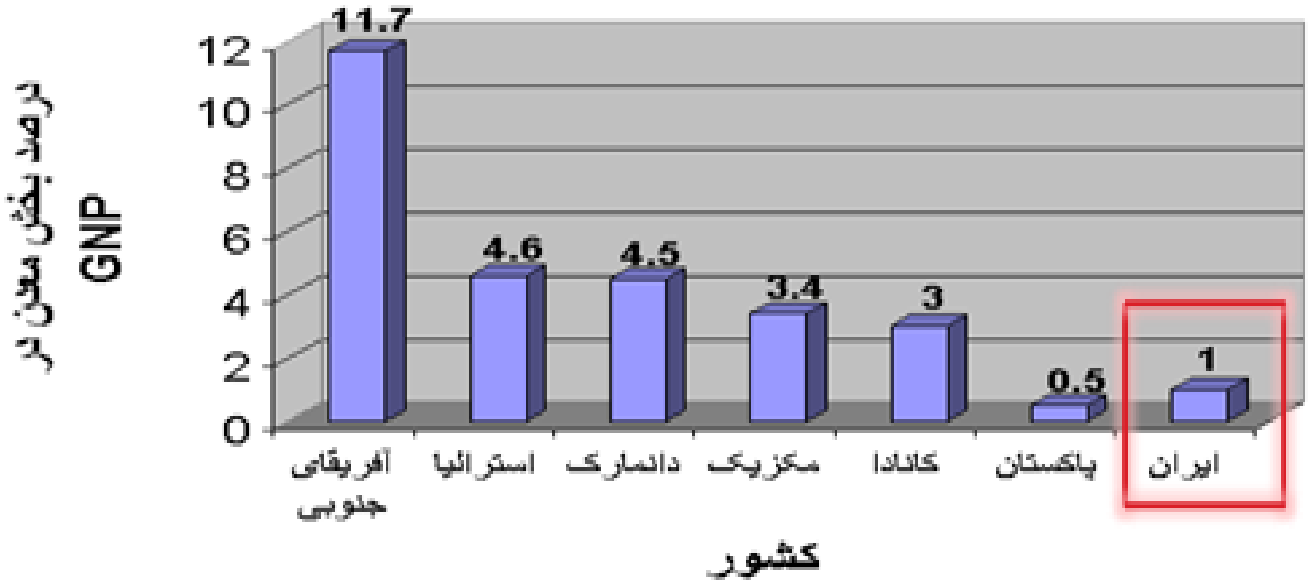

منبع: زیتون‌نژاد(۱۳۸۴)

البته روند زمانی سهم ارزش‌افزوده بخش معدن از تولید ناخالص داخلی گویای روند افزایشی آرام و تقریباً پایداری در این شاخص می‌باشد(نمودار ۳).

نمودار ۳) روند زمانی سهم ارزش‌افزوده بخش معدن از *GDP*

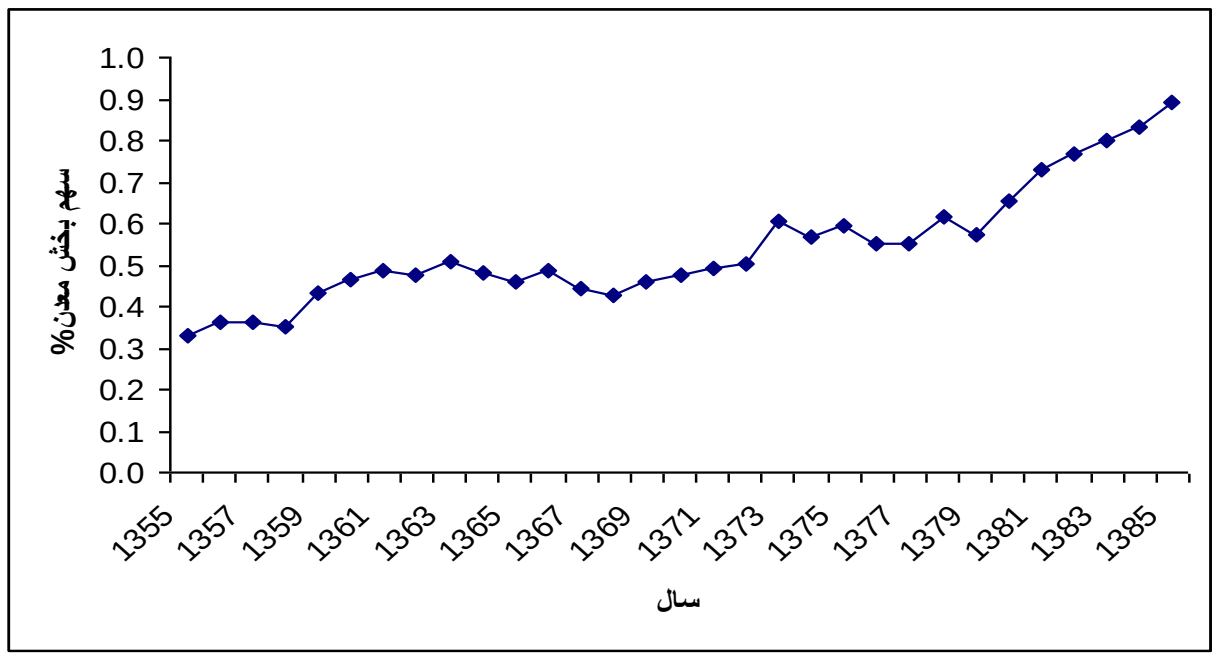

منبع: بانک مرکزی و محاسبات نویسندگان

از نظر تئوری تولید، متغیّرهای اصلی تولید عبارتند از ارزش‌افزوده(به‌عنوان ستاده)، نیروی کار و موجودی سرمایه(به‌عنوان نهاده‌ها). ارزش‌افزوده، اشتغال و موجودی سرمایهٔ بخش معدن در دورهٔ بررسی موردنظر، تقریباً یک روند افزایشی را تجربه نموده‌اند(جدول ۳).

جدول ۳) متوسط نرخ رشد مرکب سالانهٔ ارزش‌افزوده،
اشتغال و موجودی سرمایه در بخش معدن:۱۳۸۵-۱۳۵۵

| متغیّر | میانگین نرخ رشد مرکب سالیانه |
|---|---|
| ارزش‌افزوده | ٪ ۵/۴۸ |
| اشتغال نیروی‌کار | ٪ ۲/۶۵ |
| موجودی سرمایه | ٪ ۲/۲۱ |

منبع: منابع داده‌ها و محاسبات نویسندگان

## ۴. تصریح مدل و شرح متغیّرها

در این بخش ابتدا داده‌های گردآوری‌شده و منابع آنها، به‌طور کامل معرفی شده و در ادامه، مدل‌های موردنظر جهت برآورد تابع تولید و اندازه‌گیری رشد بهره‌وری کل عوامل تصریح می‌شوند.

– **ارزش‌افزوده:** طبق تعریف، ارزش‌افزوده عبارت است از مابه‌التفاوت ارزش دریافتی‌ها و ارزش پرداختی‌ها. ارزش دریافتی‌ها در واقع مجموع ارزش تولیدات مواد معدنی، ضایعات قابل‌فروش مواد معدنی، ساخت و تعمیرات اساسی اموال سرمایه‌ای توسط شاغلان معدن و سایر دریافتی‌ها می‌باشد. در مقابل، ارزش پرداختی‌ها مجموع ارزش مواد، ابزار و وسایل کار کم‌دوام، سوخت مصرف‌شده، برق خریداری‌شده، آب خریداری‌شده و سایر پرداختی‌ها می‌باشد.

– **موجودی سرمایه:** موجودی سرمایه به مفهوم مجموع کالاهای سرمایه‌ای است که برحسب یک واحد یکسان اندازه‌گیری می‌شوند. کالاهای سرمایه‌ای در بخش معدن عبارتند از ماشین‌آلات و ابزار و وسایل کار بادوام، وسایل نقلیه، لوازم و تجهیزات اداری، ساختمان و تأسیسات(بدون لحاظ ارزش زمین)، راه اختصاصی معدن، نرم‌افزارهای رایانه‌ای و ... .

– **شاغلان:** طبق تعریف، شاغلان در بخش معدن به تمام افرادی اطلاق می‌شود که در داخل یا خارج معدن به‌صورت تمام-وقت یا پاره‌وقت برای معدن کار می‌کنند. شاغلان از نظر وظیفه‌ای که به عهده دارند به دو گروه شاغلان خط تولید و شاغلان اداری، مالی و خدماتی تفکیک می‌شوند. به‌دلیل اعتبار بیشتر داده‌های سرشماری‌شده نسبت به داده‌های برآوردشده، به‌جای استفاده از برآوردهای صورت‌گرفته توسط مطالعات مختلف، از داده‌های مربوط به گزارش‌های سالانهٔ معادن کشور که هر ساله توسط مرکز آمار ایران منتشر می‌گردد، استفاده شده است. متأسفانه آمار ارائه‌شده توسط مرکز آمار ایران در مورد نیروی کار شاغل در بخش معدن در سال‌های ۱۳۶۲-۱۳۵۵، ۱۳۶۹ و ۱۳۸۳ با خلاء آماری مواجه بوده است. برای به‌دست آوردن داده-های مربوط به این سال‌ها، روش درون‌یابی برونزا بکار گرفته شده است. در این راستا از داده‌های موجود در سال‌های قبل و بعد از خلاءهای آماری به‌عنوان بنچمارک استفاده شده است.[1] بدین ترتیب، معدود خلاءهای آماری موجود در سری‌زمانی اشتغال مرکز آمار، برطرف گردیدند. جمع‌بندی متغیّرهای مورداستفاده و منابع داده‌ها به شرح جدول۴ می‌باشد.

**جدول ۴) منابع داده‌ها و تعریف متغیّرهای مورد استفاده**

| متغیّر | تعریف متغیّر | منبع |
|---|---|---|
| **Q** | ارزش افزوده(تولید)* | بانک مرکزی |
| **L** | اشتغال | مرکز آمار ایران |
| **K** | کل موجودی سرمایه* | بانک مرکزی |

* به قیمت ثابت سال ۱۳۷۶

قطعاً بسیاری از فرم‌های تابعیِ ریاضیاتی وجود دارند که قادرند با محدودسازی دنیای واقعی و پذیرش بعضی فروض، فرآیند تولید را در سطح بخش‌های اقتصادی توضیح دهند. زیتون‌نژاد(۱۳۸۷) در مطالعه‌ای، فرآیند تولید در بخش معدن

[1]. جهت دستیابی به اطلاعات بیشتر در مورد روش درونیابی برونزا ر.ک. به زیتون‌نژاد(۱۳۸۷).

ایران طی دورهٔ زمانی ۱۳۸۵-۱۳۵۵ را با استفاده از چهار نوع مختلف تابع تولید کاب- داگلاس، تابع تولید متعالی، تابع تولید دبرتین و تابع تولید ترنسلوگ مورد بررسی و برآورد قرار داده و دریافته است که از نقطه‌نظر انطباق با معیارهای آماری، اقتصادسنجی و مبانی نظری علم اقتصاد، تابع تولید کاب-داگلاس بهترین تصریح در توضیحِ فرآیند تولید بخش معدن ایران است. تابع کاب- داگلاس، یکی از توابع تولید رایج در علم اقتصاد بوده که کاربردها و محاسن فراوانی دارد. اقبال اقتصاددانان به این تابع تا حد زیادی به نتایج آماری قابل‌توجه حاصل از برازش مناسب این تابع روی داده‌های سری زمانی، ارتباط دارد(فریزر، ۲۰۰۲). این تابع در حالت وجود دو نهادهٔ تولید سنّتی نیروی کار و سرمایه، به‌قرار زیر بیان می‌شود:

$$Q = AK^{\alpha}L^{\beta} \quad [۶]$$

در این تابع، $A$ بیانگر سطح تکنولوژی، $K$ و $L$ نیز به‌ترتیب معرّفِ نهاده‌های تولیدی سرمایه و نیروی کار می‌باشند. $\alpha$ و $\beta$ نیز به ترتیب کشش تولیدیِ نهاده‌های نیروی کار و سرمایه بوده و مقادیری ثابت و مثبت می‌باشند($0 < \alpha$ و $\beta > 0$ و $A > 0$). رابطهٔ بین متغیرهای وابسته و توضیحی در تابع کاب- داگلاس، اساساً غیرخطی می‌باشد. با این وجود، در صورتی که از طرفین آن لگاریتم طبیعی بگیریم، به معادلهٔ خطی زیر دست خواهیم یافت:

$$\ln Q = \ln A + \alpha \ln K + \beta \ln L + u \quad [۷]$$

اکنون این معادله، یک مدل رگرسیون خطی بوده و این امکان را دارد که با استفاده از منطق رگرسیون خطی کلاسیک که روش حداقل مربعات معمولی را جهت برآورد به‌کار می‌گیرد، برآورد نمود. بعضاً در تابع تولید مذکور، متغیّر روند زمانی به‌منظور محاسبهٔ رشد فنّی تولید اضافه می‌گردد. این امر اولین بار توسط تینبرگن[۱] (۱۹۴۲) مطرح شد. بر اساس تعریف، رشد فنّی تولید عبارت است از درصد تغییرات در میزان ستاده، در طول زمان، با توجه به ثابت بودن عوامل تولید. به عبارت دیگر تحولات فنّی تولید، بیان‌کنندهٔ درصد تغییرات در محصول به‌ازای درصد تغییرات در زمانند. بنابراین، تحولات فنّی نشانگر تأثیر زمان بر استفادهٔ مطلوب‌تر و بهتر از عوامل تولید(با فرض ثابت بودن همهٔ عوامل تولید) هستند. بنابراین این معیار، میزان انتقال تابع تولید را توضیح می‌دهد.

اگر $\alpha + \beta = 1$ یا به عبارتی دیگر $\beta = 1 - \alpha$ باشد، در این حالت با فرم مقیّد تابع کاب- داگلاس مواجه خواهیم بود. در صورتی که از طرفین آن لگاریتم طبیعی بگیریم، به معادلهٔ خطی زیر دست خواهیم یافت:

$$\ln \frac{Q}{L} = \ln A + \alpha \ln \frac{K}{L} + u \quad [۸]$$

باید توجه داشت که به علت اجتناب از مشکل هم‌خطی بین $\ln K$ و $\ln L$ ممکن است خطای معیار کاهش یابد. بدیهی است که برای استفاده از رویکرد حسابداری رشد (در اندازه‌گیری سهم رشد بهره‌وری از رشد اقتصادی)، تابع تولید کاب-داگلاس سرانه مقیّد را برآورد خواهد شد. قبل از برآورد این تابع، فرض بازدهی ثابت نسبت به مقیاس، از طریق تصریح تابع کاب- داگلاس غیرمقیّد و آزمون والد[۲]، آزمون می‌شود.

---

[۱]. Tinbergen
[2]. Wald test

برای اندازه‌گیری رشد بهره‌وری کل عوامل، از روش باقیماندۀ سولو، که در حقیقت چیزی جز تفاضل میانگین موزون رشد عوامل، از رشد تولید نیست، استفاده می‌شود. این موضوع به زبان سادۀ ریاضی به صورت زیر قابل بیان است:

$$TF\dot{P} = \dot{V} - \alpha\dot{K} - \beta\dot{L} \quad [۹]$$

## ۵. نتایج تجربی

در این بخش، ابتدا فرضیۀ بازدهی ثابت نسبت به مقیاس در بخش معدن ایران از طریق تابع تولید کاب- داگلاس غیرمقیّد بخش معدن، مورد آزمون قرار می‌گیرد. در ادامه، با استفاده از داده‌های سری‌زمانی ۱۳۸۵-۱۳۵۵، تابع تولید مقیّد بخش معدن در اقتصاد ایران برآورد می‌گردد تا از آن طریق و با در نظر گرفتن فرض شرایط رقابت کامل، بتوان سهم درآمدی هر یک از عوامل تولید را محاسبه نمود. سپس با مدنظر قرار دادن رویکرد حسابداری رشد و با استفاده از پارامترهای برآورد-شده، رشد بهره‌وری کل عوامل در بخش معدن ایران اندازه‌گیری می‌گردد.

برای استفاده از روش حسابداری رشد، لازم است تابع تولید مقیّد برآورد شود. زیتون نژاد(۱۳۸۷ و ۲۰۱۵) تابع تولید کاب- داگلاس غیرمقیّد بخش معدن ایران را برای دورۀ زمانی ۱۳۸۵-۱۳۵۵ برآورد نموده است. نتایج این برآورد نشان می-دهد که کشش تولید نسبت به نهاده‌های سرمایه و نیروی کار در شرایط غیرمقیّد، به‌ترتیب برابر ۰/۴۴ و ۰/۴۱ بوده است. وی در ادامه و از طریق آزمون والد نشان داده است که فرضیۀ وجود بازدهی ثابت نسبت به مقیاس($\alpha+\beta=1$) در بخش معدن ایران رد نمی‌شود. بنابراین در ادامه، با استفاده از نرم‌افزار *Eviws6*، تابع تولید مقیّد بخش معدن برآورد می‌شود، تا بر اساس آن بتوان محاسبات مربوط به رویکرد حسابداری رشد را انجام داد. قبل از برآورد تابع تولید مقیّد، نخستین گام، انجام آزمون پایایی متغیرها است. یک متغیّر سری‌زمانی هنگامی پایاست که میانگین، واریانس و ضرایب خودهمبستگی آن در طول زمان ثابت باقی بماند. آزمون‌های ریشه واحد نظیر دیکی- فولر تعمیم‌یافته[1] و فیلیپس- پرون[2] از معمول‌ترین آزمون‌هایی است که برای تشخیص پایایی یک فرآیند سری‌زمانی مورد استفاده قرار گرفته و آماره حاصله با $t$ مکینون(مقادیر بحرانی) مقایسه می-گردد. نتایج آزمون‌های پایایی برای متغیرهای ارزش‌افزوده سرانه و سرمایه سرانه به‌طور خلاصه در جدول ۷ ارائه می‌گردد.

**جدول۵) خلاصۀ نتایج آزمون‌های پایایی انجام‌شده برای لگاریتم متغیرهای سرانه**

| نام متغیر | سطح ADF | سطح PP | تفاضل مرتبۀ اول ADF | تفاضل مرتبۀ اول PP | نتیجه |
|---|---|---|---|---|---|
| LKBARL | -۱/۶۹۳ | -۱/۸۰٤ | -٥/۰۰٥ | -٥/۰۰۲ | I(1) |
| LQBARL | -۱/۰۳٤ | -۱/۰٤۰ | -٤/۷٦٤ | -٤/۷۲٤ | I(1) |

*. در آزمون سطح، مقادیر بحرانی در سطوح ۱٪، ۵٪ و ۱۰٪ به ترتیب برابر (-٤/۲۹۷)، (-۳/٥٦۸) و (-۳/۲۱۸) و برای آزمون تفاضل مرتبۀ اول، مقادیر بحرانی در سطوح ۱٪، ۵٪ و ۱۰٪ به ترتیب برابر (-٤/۳۱۰)، (-۳/٥۷٤) و (-۳/۲۲۲) می‌باشد.

منبع: یافته‌های نویسندگان

1. Augmented Dickey Fuller
2. Philips-Perron

همان‌طور که مشاهده می‌گردد، نتایج حاصل از هر دو آزمون پایایی برای هر دو متغیر در سطح، بیانگر وجود ریشهٔ واحد و ناپایا بودن آن‌ها می‌باشد. ولیکن نتایج آزمون هر دو متغیر در تفاضل مرتبهٔ اول، بیانگر I(1) بودن متغیرها و به بیان دیگر، عدم وجود ریشهٔ واحد و در نتیجه پایابودن تفاضل مرتبهٔ اول آن‌ها می‌باشد. اکنون، شرایط استفاده از روش هم‌جمعی مهیاست. حال پس از اتمام مرحلهٔ اول به مرحلهٔ دوم از مراحل چهارگانهٔ روش هم‌جمعی خواهیم پرداخت. در این راستا، تابع تولید کاب- داگلاس مقیّد برآورد می‌گردد(جدول ۶).

**جدول٦) خلاصهٔ نتایج حاصل از برآورد تابع تولید مقیّد بخش معدن**

| پارامترها، آماره‌ها و معیارها | مقادیر عددی | پارامترها، آماره‌ها و معیارها | مقادیر عددی |
|---|---|---|---|
| کشش تولید نسبت به نهادهٔ سرمایه( $\alpha$ ) | ۰/۵۲ | $R^2$ | ٪۹٤ |
| t-Statistic مربوط به $\alpha$ | ۲/۸٦ | $\bar{R}^2$ | ٪۹۳ |
| کشش تولید نسبت به نهادهٔ نیروی‌کار( $\beta$ ) | ۰/٤۸ | F-Statistic | ۲۰٦ |
| t-Statistic مربوط به $\beta$ | ۲/۸٦ | دوربین- واتسون(D.W) | ۱/۷۹ |
| ضریب AR(1) | ۱/۰۸ | تعداد مشاهدات(n) بعد از تعدیل | ۳۰ |
| t-Statistic مربوط به ضریب AR(1) | ۱۹/۳۸ | سطح اطمینان ضرایب | ٪۹۹ |

نتایج حاصل از این برآورد نشان می‌دهد که کشش تولید نسبت به نهاده‌های سرمایه و نیروی کار(یا با فرض رقابت کامل، همان سهم نهاده‌ها از ارزش‌افزوده) به‌ترتیب برابر ۰/۵۲ و ۰/۴۸ بوده است. نکتهٔ جالب در هر دو برآورد تابع تولید(مقیّد و غیرمقیّد) ثابت‌بودن نسبت کشش سرمایه به کشش نیروی کار ( $\frac{\alpha}{\beta}$ ) بوده است. این مقدار در تابع غیرمقیّد برابر ۱/۰۸ و در تابع تولید مقیّد برابر ۱/۰۷ بوده است.[1] این مسئله تا حدی بیانگر پایداری مدل‌های برآوردی نیز می‌باشد. برای رفع مشکل خودهمبستگی در مدل فوق با استفاده از آزمون فرآیند باکس-جنکینز[2]، جزء AR(1) به مدل افزوده شده است. آمارهٔ دوربین- واتسون حدود ۱/۸ بوده و بیانگر این است که مشکل خودهمبستگی در مدل برآوردشده وجود ندارد. ضریب تبیین و ضریب تبیین تعدیل‌شده در مدل فوق به‌ترتیب برابر ۹۴٪ و ۹۳٪ درصد می‌باشد که بیانگر توضیح‌دهی بسیار مناسب مدل فوق می‌باشد. در مدل مقیّد برآوردی، آمارهٔ معناداری ضرایب(*t-statistic*) نیز دارای وضعیت مناسبی است، به‌طوری که تمام ضرایب متغیرهای توضیحی، در سطح اطمینان ۹۹٪ مورد تأیید و معنادار هستند. آمارهٔ $F$ نیز بیانگر معناداری کل الگو می‌باشد($F$=۲۰۶). بنابراین، مدل برآوردشده با توجه به معیارهای آماری، تئوریک و اقتصادسنجی از شاخص‌های قابل‌قبولی برخوردار بوده و بدین جهت به‌عنوان یک الگو، مورد قبول واقع شده است. اکنون، الگو در راستای بررسی برقراری و یا نقض فروض کلاسیک مانند همسانی واریانس، عدم خودهمبستگی، نُرمالیتی و ... مورد ارزیابی قرار می‌گیرد. بدین ترتیب درجهٔ اطمینان و اعتبار الگو مشخص خواهد شد.

---

[1]. زیتون نژاد (۱۳۸۷ و ۲۰۱۵) تابع تولید بخش معدن را با تصریح های مختلف از جمله تابع کاب داگلاس غیرمقید برآورد نموده است و نسبت ضرایب برآوردی در آن مطالعه برابر با ۱/۰۸ بوده است.

[2]. Box-Jenkins

**آزمون‌ها و بررسی فروض**

یکی از فروض کلاسیک، یکسان بودن واریانس اجزای اخلال در دوره‌های مختلف است. با انجام آزمون ناهمسانی واریانس رگرسیون از طریق روش بریوش-پاگان-گادفری[۱]، مشخص گردید که فرضیهٔ وجود ناهمسانی واریانس در برآورد انجام‌شده، رد می‌گردد. جهت بررسی مشکل خودهمبستگی، از آزمون $LM$[۲] استفاده شده است. نتایج حاصل از انجام این آزمون برای مدل غیرمقیّد اولیه(قبل از وارد نمودن جزء $\mathrm{AR}(1)$)، حاکی از وجود مشکل خودهمبستگی در مدل برآوردی اولیه بوده است(نمودار ۴، سمت چپ). لذا جهت رفع این مشکل، جزء $\mathrm{AR}(1)$ به‌مدل افزوده شد، و بدین‌ترتیب این مشکل، در مدل مقیّد نهایی رفع گردید(نمودار ۴، سمت راست).

**نمودار ٤) وضعیت خودهمبستگی و همبستگی جزئی اجزاءاخلال در مدل مقیّد اولیه(سمت چپ) و نهایی(سمت راست)**

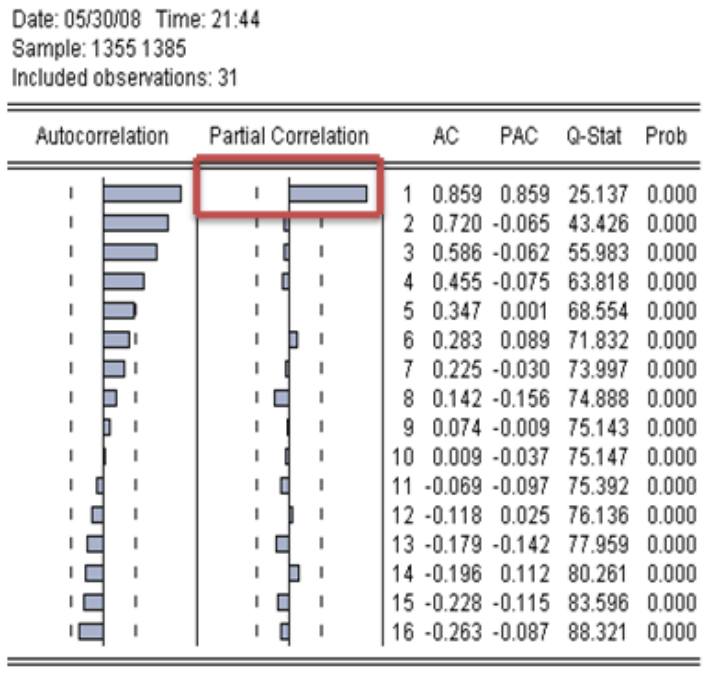

Date: 05/30/08 Time: 21:44
Sample: 1355 1385
Included observations: 31

| | AC | PAC | Q-Stat | Prob |
|---|---|---|---|---|
| 1 | 0.859 | 0.859 | 25.137 | 0.000 |
| 2 | 0.720 | -0.065 | 43.426 | 0.000 |
| 3 | 0.586 | -0.062 | 55.983 | 0.000 |
| 4 | 0.455 | -0.075 | 63.818 | 0.000 |
| 5 | 0.347 | 0.001 | 68.554 | 0.000 |
| 6 | 0.283 | 0.089 | 71.832 | 0.000 |
| 7 | 0.225 | -0.030 | 73.997 | 0.000 |
| 8 | 0.142 | -0.156 | 74.888 | 0.000 |
| 9 | 0.074 | -0.009 | 75.143 | 0.000 |
| 10 | 0.009 | -0.037 | 75.147 | 0.000 |
| 11 | -0.069 | -0.097 | 75.392 | 0.000 |
| 12 | -0.118 | 0.025 | 76.136 | 0.000 |
| 13 | -0.179 | -0.142 | 77.959 | 0.000 |
| 14 | -0.196 | 0.112 | 80.261 | 0.000 |
| 15 | -0.228 | -0.115 | 83.596 | 0.000 |
| 16 | -0.263 | -0.087 | 88.321 | 0.000 |

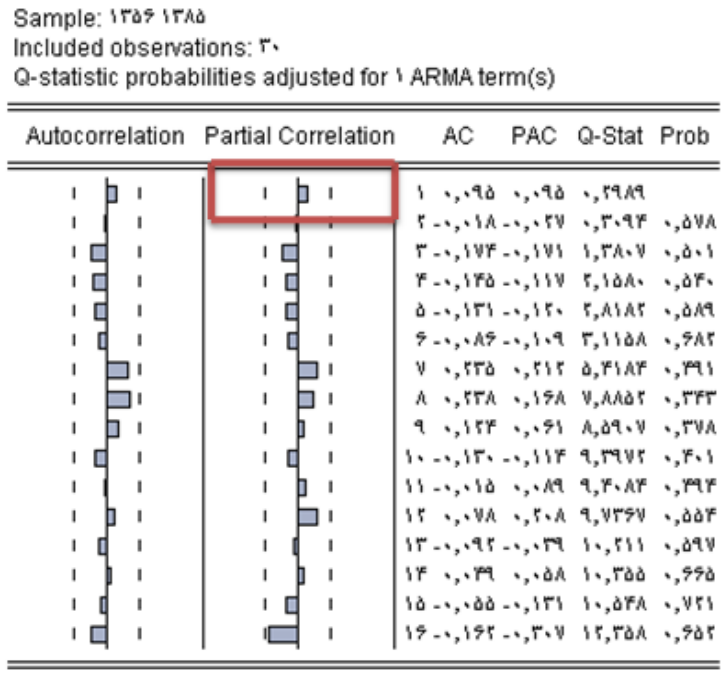

Sample: ۱۳۵۶ ۱۳۸۵
Included observations: ۳۰
Q-statistic probabilities adjusted for ۱ ARMA term(s)

| | AC | PAC | Q-Stat | Prob |
|---|---|---|---|---|
| ۱ | ۰,۰۹۵ | ۰,۰۹۵ | ۰,۲۹۸۹ | |
| ۲ | -۰,۰۱۸ | -۰,۰۲۷ | ۰,۳۰۹۴ | ۰,۵۷۸ |
| ۳ | -۰,۱۷۴ | -۰,۱۷۱ | ۱,۳۸۰۷ | ۰,۵۰۱ |
| ۴ | -۰,۱۴۵ | -۰,۱۱۷ | ۲,۱۵۸۰ | ۰,۵۴۰ |
| ۵ | -۰,۱۳۱ | -۰,۱۲۰ | ۲,۸۱۸۲ | ۰,۵۸۹ |
| ۶ | -۰,۰۸۶ | -۰,۱۰۹ | ۳,۱۱۵۸ | ۰,۶۸۲ |
| ۷ | ۰,۲۳۵ | ۰,۲۱۲ | ۵,۴۱۸۴ | ۰,۴۹۱ |
| ۸ | ۰,۲۳۸ | ۰,۱۶۸ | ۷,۸۸۵۲ | ۰,۳۴۳ |
| ۹ | ۰,۱۲۴ | ۰,۰۶۱ | ۸,۵۹۰۷ | ۰,۳۷۸ |
| ۱۰ | -۰,۱۳۰ | -۰,۱۱۴ | ۹,۳۹۷۲ | ۰,۴۰۱ |
| ۱۱ | -۰,۰۱۵ | ۰,۰۸۹ | ۹,۴۰۸۴ | ۰,۴۹۴ |
| ۱۲ | ۰,۰۷۸ | ۰,۲۰۸ | ۹,۷۳۶۷ | ۰,۵۵۴ |
| ۱۳ | -۰,۰۹۲ | -۰,۰۳۹ | ۱۰,۲۱۱ | ۰,۵۹۷ |
| ۱۴ | ۰,۰۳۹ | ۰,۰۵۸ | ۱۰,۳۵۵ | ۰,۶۶۵ |
| ۱۵ | -۰,۰۵۵ | -۰,۱۳۱ | ۱۰,۵۴۸ | ۰,۷۲۱ |
| ۱۶ | -۰,۱۶۲ | -۰,۳۰۷ | ۱۲,۳۵۸ | ۰,۶۵۲ |

نتایج حاصل از بررسی نرمال بودن توزیع اجزاءاخلال حاکی از نرمال بودن توزیع آن‌ها می‌باشد(نمودار ۵).

**نمودار۵) توزیع آماری اجزاءاخلال در مدل مقیّد**

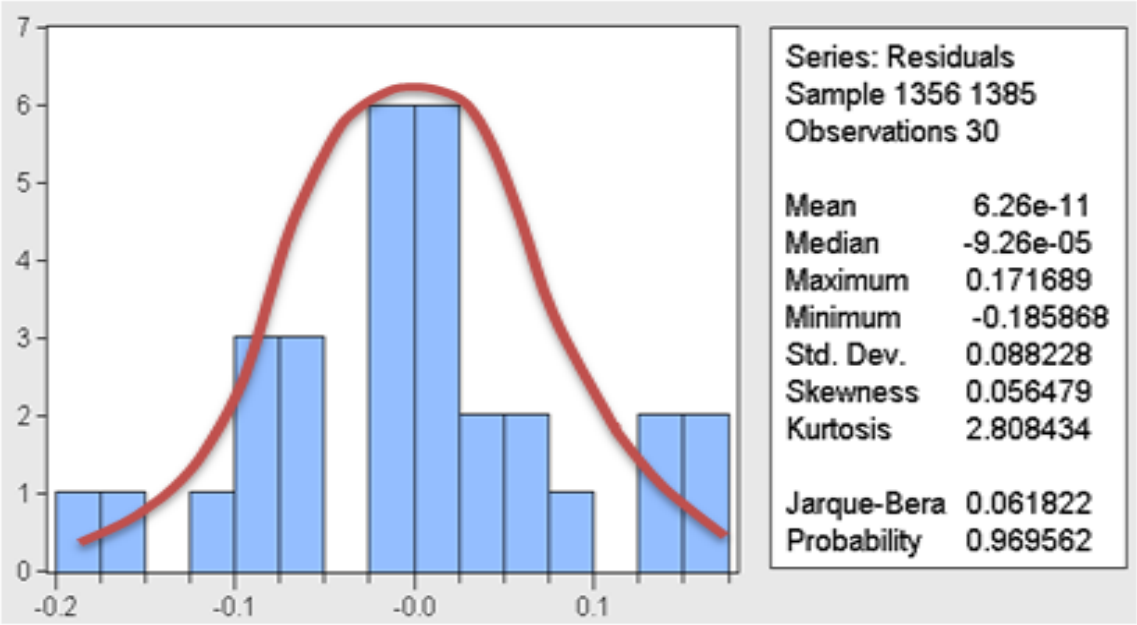

بنابراین، نمودار بالا و سایر نتایج از قبیل ضریب جارک-برا[۳] نشان می‌دهد که فرض توزیع نرمال برای فرآیند تصادفی جزءاخلال برقرار بوده و به عبارت دیگر شاهد وضعیت $u \sim N(0,\sigma^2)$ می‌باشیم. این قضیه به‌نوبهٔ خود مبیّن $E(u_i)=0$

[1]. Breusch-Pagan-Godfrey
[2]. LM test
[3]. Jarque-Bera

است که نشانگر برقراری یکی دیگر از فروض کلاسیک می‌باشد. ضمناً محاسبهٔ $E(u_i)$ نشان می‌دهد که مقدار عددی آن برابر ۱۱-۶/۲۶e است که تقریباً معادل عدد صفر می‌باشد. پنجمین فرض از فروض کلاسیک، فرض مستقل‌بودن اجزاءاخلال از متغیرهای توضیحی می‌باشد. نتایج حاصل از محاسبهٔ ضریب همبستگی بین اجزاءاخلال و متغیر توضیحی لگاریتم سرمایهٔ سرانه معادل ۰/۱۱- بوده است که بیانگر مستقل بودن اجزاءاخلال از هر دو متغیر توضیحی در مدل برآوردشده می‌باشد. لذا فرض استقلال اجزاءاخلال از متغیر توضیحی مدل نیز برقرار است. بنابراین، مشخص گردید که در مدل برآوردشده، تمام فروض کلاسیک برقرار بوده و مشکلی به‌عنوان نقض فروض کلاسیک وجود نداشته است. ضمناً از آنجا که مدل مقیّد برآوردشده، صرفاً دارای یک متغیر توضیحی است، لذا احتمال بروز هرگونه مشکل هم‌خطی برای مدل فوق، غیرممکن می‌باشد.

اکنون، به‌عنوان مراحل سوم و چهارم از روش هم‌جمعی، لازم است که سری‌زمانی اجزاءاخلال را محاسبه نموده و پایایی این سری را مورد آزمون قرار داد. بدیهی است که تنها در صورتی که این سری‌زمانی در سطح پایا باشد، نتایج حاصل از برآورد حداقل مربعات معمولی که محاسبه شده است، اعتبار خواهد داشت. سری اجزاءاخلال با استفاده از خروجی‌های نرم‌افزار *Eviws6* به‌دست آمده و خلاصه نتایج آزمون‌های پایایی *ADF* و *PP* به شرح جدول زیر می‌باشد.

**جدول ۷) خلاصهٔ نتایج آزمون‌های پایایی انجام‌شده برای اجزاءاخلال مدل مقیّد**

| نام متغیر | سطح ADF | سطح PP | نتیجه |
|---|---|---|---|
| اجزاءاخلال | ٤/۹۹۹- | ٦/۰۱۹- | I(0) |

*. در آزمون سطح، مقادیر بحرانی در سطوح ۱٪، ۵٪ و ۱۰٪ به ترتیب برابر (٤/۳۱۰-)، (۳/۵۷٤-) و (۳/۲۲۲-) می‌باشد. منبع: یافته‌های نویسندگان

نتایج حاصل از هر دو آزمون ریشهٔ واحد برای سری اجزاءاخلال نشان می‌دهد که این سری در سطح پایا است. بنابراین، مدل برآوردشده با توجه به معیارهای آماری، تئوریک و اقتصادسنجی از شاخص‌های قابل قبولی برخوردار است.

**اندازه‌گیری رشد بهره‌وری کل عوامل**

اکنون، با استفاده از روش باقیماندهٔ سولو(معادلهٔ [۹]) و با در نظر گرفتن نتایج حاصل از تابع تولید کاب- داگلاس مقیّد، رشد بهره‌وری کل عوامل برای هر سال و متوسط نرخ رشد سالانهٔ بهره‌وری کل عوامل در طول دورهٔ مورد مطالعه، اندازه‌گیری می‌شود(جدول ۸).

جدول ۸) نرخ رشد بهره‌وری کل عوامل در بخش معدن

| سال | نرخ رشد بهره‌وری کل عوامل | سال | نرخ رشد بهره‌وری کل عوامل | سال | نرخ رشد بهره‌وری کل عوامل |
|---|---|---|---|---|---|
| ۱۳۵۶ | -٪۳/٤٦ | ۱۳٦٦ | ۱/٤٤ | ۱۳۷٦ | -٪٦/۰۰ |
| ۱۳۵۷ | -٪۱۳/۸۸ | ۱۳٦۷ | -٪۵/۵۵ | ۱۳۷۷ | ٪۸/٦۲ |
| ۱۳۵۸ | -٪۱۰/۵۰ | ۱۳٦۸ | -٪۰/٤۸ | ۱۳۷۸ | ٪۱۲/۹۷ |
| ۱۳۵۹ | ٪۱/٦۸ | ۱۳٦۹ | ٪۱۷/۳٦ | ۱۳۷۹ | -٪۲/۸۲ |
| ۱۳٦۰ | ٪۱/۵۲ | ۱۳۷۰ | ٪۰/۹۵ | ۱۳۸۰ | ٪٦/۳۳ |
| ۱۳٦۱ | ٪۱٤/٤۲ | ۱۳۷۱ | ٪۱/۹۹ | ۱۳۸۱ | ٪۱۹/۰۷ |
| ۱۳٦۲ | ٪۵/٦۷ | ۱۳۷۲ | -٪٦/۳۱ | ۱۳۸۲ | ٪۱۱/۳۸ |
| ۱۳٦۳ | ٪۱/۲۰ | ۱۳۷۳ | ٪۱۸/۰۵ | ۱۳۸۳ | ٪۵/۵۸ |
| ۱۳٦٤ | -٪۵/۰۹ | ۱۳۷٤ | ٪۲/۵۱ | ۱۳۸٤ | ٪۱۰/۷۰ |
| ۱۳٦۵ | -٪۱٦/٦۷ | ۱۳۷۵ | ٪۹/۲۵ | ۱۳۸۵ | ٪۸/۳۳ |
| | | | میانگین کل دوره = ٪۲/۹٤ | | |

منبع: برآوردها و محاسبات نویسندگان

میانگین نرخ رشد بهره‌وری کل عوامل طی دورۀ ۱۳۸۵-۱۳۵۵ معادل ٪۲/۹۴ است. این موضوع بیانگر آن است که متوسط سطح بهره‌وری کل عوامل طی دورۀ مذکور در حال افزایش بوده است. در ادامه، به‌منظور شناخت بیشتر وضعیت رشد بهره‌وری در بخش معدن، نمودار روند زمانی رشد بهره‌وری در بخش معدن، ترسیم می‌گردد(نمودار ۶).

نمودار ۶) نرخ رشد بهره‌وری کل عوامل در بخش معدن طی ۱۳۸۵-۱۳۵۶

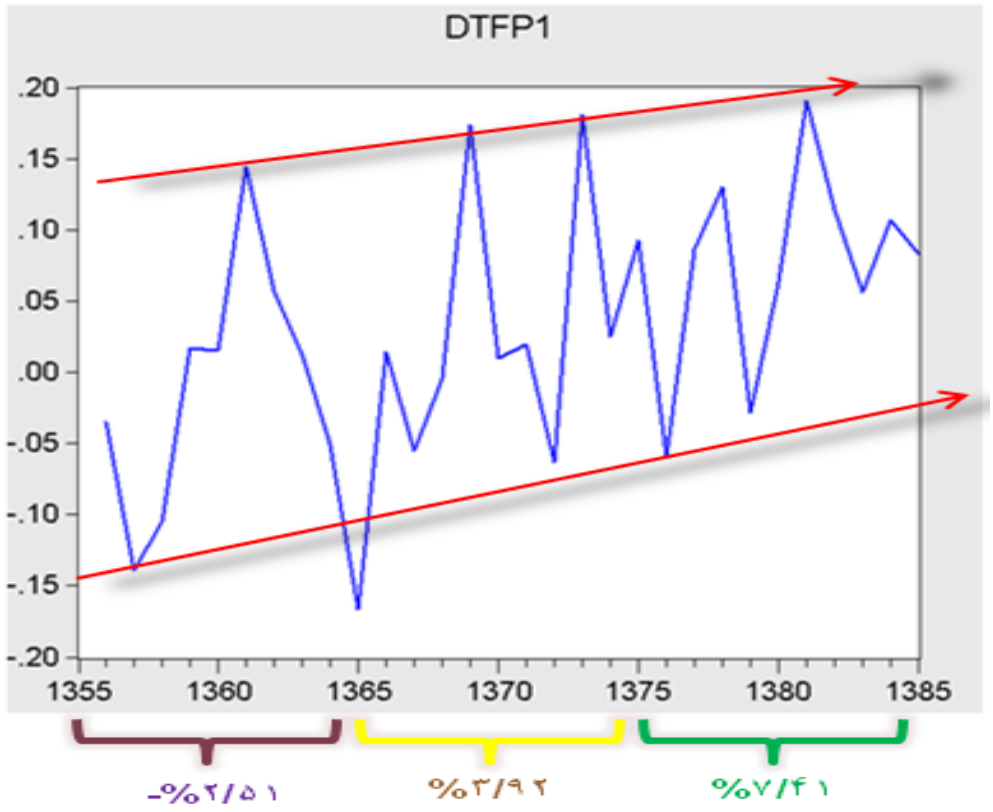

متوسط نرخ رشد سالیانۀ ۲/۹۴ درصدی بهره‌وری کل عوامل، نشانگر این بود که متوسط سطح بهره‌وری کل عوامل طی دورۀ بررسی، در حال افزایش بوده است. از طرف دیگر، روند افزایشی نمودار فوق، بیانگر افزایش میانگین این نرخ رشد در طول زمان است. به عبارت دیگر، بخش معدن در سه دهۀ اخیر به مرور، روند افزایشی شتابان‌تر و یا به عبارتی رشد سریع‌تری

را در بهره‌وری تجربه نموده است. تداوم این روند نویددهندهٔ آینده‌ای نزدیک‌تر برای نیل به سطوح بالا و بالاتری از بهره‌وری در بخش معدن خواهد بود.

برای بررسی دقیق‌تر این موضوع، دورهٔ زمانی تحقیق به پنج زیربازهٔ زمانی ۱۳۶۷-۱۳۵۶(قبل از تدوین برنامه‌های توسعهٔ اقتصادی)، ۱۳۷۳-۱۳۶۸(برنامهٔ اول و سال ۱۳۷۳)، ۱۳۷۸-۱۳۷۴(برنامهٔ دوم)، ۱۳۸۳-۱۳۷۹(برنامهٔ سوم) و ۱۳۸۵-۱۳۸۴(دو سال اول برنامهٔ چهارم) تقسیم می‌شود. نمودار زیر متوسط نرخ رشد سالیانهٔ بهره‌وری کل عوامل را، به‌تفکیک، برای هر بازهٔ زمانی ارائه می‌کند.

نمودار ۷) متوسط رشد بهره‌وری کل عوامل در بخش معدن طی دوره‌های منتخب

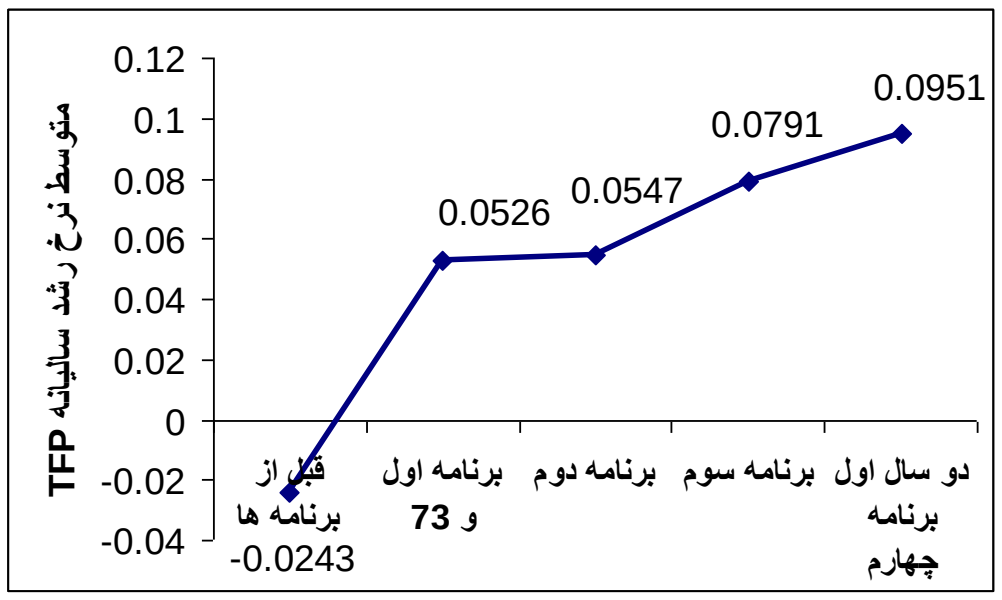

منبع: برآوردها و محاسبات نویسندگان

نمودار فوق نیز به نوعی مؤیّد یافته‌های حاصل از نمودار ۶ است. همان‌طور که در نمودار ۱۵ نیز مشخص بود، متوسط نرخ رشد سالیانهٔ $TFP$ در طول زمان در حال افزایش بوده که نشان‌دهندهٔ روند رو به رشد مناسب بهره‌وری در بخش معدن می‌باشد.

**منابع رشد اقتصادی بخش معدن**

برای محاسبهٔ سهم عوامل رشد در رشد ستاده، به اطلاعات رشد سالانهٔ متغیرهای ارزش‌افزوده، نیروی کار و سرمایه، نیاز می‌باشد. این اطلاعات در جدول ۳ گزارش شده‌اند. بر این اساس، سهم منابع رشد اقتصادی بخش معدن ایران طی دورهٔ ۱۳۵۵ تا ۱۳۸۵به شرح جدول ۹ بوده است.

**جدول ۹) سهم انباشت عوامل تولید و رشد بهره‌وری کل عوامل در رشد اقتصادی بخش معدن**

| عامل | میزان واحد درصد(٪) | بر حسب درصد |
|---|---|---|
| انباشت سرمایه | ۱/۱۵ | ٪۲۰/۹۸ |
| انباشت نیروی کار | ۱/۲۷ | ٪۲۳/۱۸ |
| رشد بهره‌وری کل‌عوامل | ۳/۰۶ | ٪۵۵/۸٤ |
| جمع | ۵/٤۸ | ٪۱۰۰ |

منبع: برآوردها و محاسبات نویسندگان

مشخص است که میانگین سهم رشد بهره‌وری کل عوامل از رشد اقتصادی بخش معدن طی دورهٔ بررسی، ٪۵۵/۸۴ می‌باشد. همچنین نتایج نشان‌دهندهٔ سهم بیشتر رشد بهره‌وری در رشد ارزش‌افزوده بخش معدن نسبت به سایر عوامل در دورهٔ

موردبررسی است. در ادامه، با تفکیک دورهٔ بررسی به پنج بازهٔ زمانی مختلف، به بررسی دقیق‌تر سهم عوامل رشد در رشد اقتصادی بخش معدن طی بازه‌های زمانی فوق، خواهیم پرداخت.

جدول ۱۰) تفکیک سهم عوامل رشد از رشد اقتصادی بخش معدن (حسابداری رشد در بخش معدن)

| بازهٔ زمانی | متوسط نرخ رشد سالانهٔ نیروی کار | متوسط نرخ رشد سالانهٔ سرمایه | متوسط نرخ رشد سالانهٔ ارزش‌افزوده | متوسط نرخ رشد سالانهٔ *TFP* | سهم افزایش نیروی کار از رشد ستاده * | سهم انباشت سرمایه از رشد ستاده * | سهم رشد بهره-وری کل از رشد ستاده * |
|---|---|---|---|---|---|---|---|
| ۱۳۶۷-۱۳۵۶ (قبل از برنامه‌ها) | ٪۳/۸۰ | ٪۱/۲۶ | ٪۰/۰۱ | -٪۲/۴۳ | ۱/۸۲۴۰ | ۰/۶۵۵۲ | -۲/۴۷۹۱ |
| ۱۳۷۳-۱۳۶۸ (برنامهٔ اول و سال۱۳۷۳) | ٪۴/۱۱ | ٪۹/۴۳ | ٪۱۳/۹۱ | ٪۵/۲۶ | ۱/۹۷۲۸ | ۴/۹۰۶۳ | ۷/۰۳۰۹ |
| ۱۳۷۸-۱۳۷۴ (برنامهٔ دوم) | ٪۲/۶۵ | -٪۳/۹۹ | ٪۵/۴۹ | ٪۵/۴۷ | ۱/۲۷۲۰ | -۲/۰۷۴۸ | ۶/۲۹۲۸ |
| ۱۳۸۳-۱۳۷۹ (برنامهٔ سوم) | ٪۰/۳۸ | ٪۶/۰۱ | ٪۱۴/۸۰ | ٪۷/۹۱ | ۰/۱۸۲۴ | ۳/۱۲۵۲ | ۱۱/۴۹۲۴ |
| ۱۳۸۵-۱۳۸۴ (دو سال اول برنامهٔچهارم) | ٪۹/۴۹ | -٪۰/۲۳ | ٪۱۳/۳۸ | ٪۹/۵۱ | ۴/۵۵۵۲ | ۰/۱۱۹۶ | ۸/۷۰۵۲ |

*. برحسب "واحد درصد"        منبع: برآوردها و محاسبات نویسندگان

همان‌طور که از اطلاعات جدول فوق برمی‌آید، طی دورهٔ بررسی، بهره‌وری کل عوامل و اشتغال به‌ترتیب دارای رشدهای پایدارتری نسبت به موجودی سرمایه بوده‌اند. رشد پایدار در بهره‌وری کل عوامل و اشتغال تا حد زیادی به رشد پایدار اقتصادی در بخش معدن کمک نموده است. در مقابل، نوسانات موجود در رشد موجودی سرمایه، که نوعاً ناشی از نوسانات موجود در سرمایه‌گذاری می‌باشند، در جهت مخالف عمل نموده است. با این همه، به علت سهم و نقش مؤثرتر رشد تقریباً پایدار بهره‌وری کل عوامل، مانع از بروز اثرات رشد نوسانی سرمایه، بر رشد ستادهٔ بخش معدن شده است.

**نتیجه‌گیری و پیشنهادات**

هدف این مقاله بررسی منابع رشد اقتصادی بخش معدن در دورهٔ زمانی ۱۳۸۵-۱۳۵۵ بود که با استفاده از برآورد تابع تولید مقیّد و محاسبهٔ رشد بهره‌وری کل عوامل، نسبت به این امر اقدام شد. نتایج نشان می‌دهد که کشش تولید نسبت به نهاده‌های سرمایه و نیروی کار(یا با فرض رقابت کامل، سهم نهاده‌ها از تولید) در شرایط مقیّد به‌ترتیب برابر ۰/۵۲ و ۰/۴۸ بوده است. از طرف دیگر، متوسط نرخ رشد سالیانهٔ ۲/۹۴ درصدی بهره‌وری کل عوامل، نشانگر این بود که متوسط سطح بهره-وری کل عوامل طی دورهٔ بررسی، در حال افزایش بوده است. ضمناً، روند افزایشی این متغیر، بیانگر افزایش میانگین این رشد در طول زمان است. با انجام محاسبات حسابداری رشد، مشخص گردید که میانگین سهم رشد بهره‌وری کل عوامل، رشد نیروی کار و رشد سرمایه از رشد اقتصادی بخش معدن طی دورهٔ بررسی، به‌ترتیب برابر ٪۵۶ ، ٪۲۳ و ٪۲۱ بوده است. همچنین نتایج نشان‌دهندهٔ سهم بیشتر رشد بهره‌وری در رشد ارزش‌افزوده بخش معدن نسبت به سایر عوامل در دورهٔ موردبررسی است. از این امر می‌توان نتیجه گرفت که طی دورهٔ بررسی، سیاست بهره‌گیری از منابع موجود در بخش معدن، در کنار سیاست رشد عوامل تولید، موجب رشد ارزش‌افزودهٔ این بخش گردیده است. با این حال، طی این دوره، سیاست رشد بهره‌وری با جدیّت بیشتری نسبت به سیاست رشد عوامل تولید، دنبال شده است. براساس یافته‌های این پژوهش و سایر پژوهش‌های مرتبط که در فصل ادبیات تجربی بررسی شد، توصیه‌های کلی در دو سطح متفاوت ارائه می‌گردد(جدول ۱۱).

جدول ۱۱) پیشنهادات و توصیه‌های کلی

| توصیه‌هایی در سطح کلان: | توصیه‌هایی در سطح بخش و بنگاه: |
| --- | --- |
| • فراهم نمودن بسترهای لازم برای توسعهٔ سرمایه‌گذاری در بخش معدن.[1]<br>• گسترش شرایط رقابت‌پذیری اقتصاد از راه‌هایی نظیر افزایش درجهٔ باز بودن اقتصاد و خصوصی‌سازی.[2]<br>• ایجاد ثبات در متغیرهای کلان اقتصادی از قبیل تورم، رشد اقتصادی و تأمین امنیت سرمایه‌گذاری با هدف ایجاد بسترسازی برای جذب سرمایه‌گذاری داخلی و خارجی.<br>• توسعهٔ زیرساخت‌ها با هدف افزایش انگیزه بخش خصوصی برای تولید مواد معدنی.<br>• اصلاح قوانین(قانون کار، قانون معادن، قانون تجارت، قانون صادرات و ...) در راستای افزایش سطح بهره‌وری و کارایی اقتصادی، به‌طوری که نظام‌های انگیزشی در آن‌ها لحاظ شده باشد.<br>• کاهش حجم دولت در فعالیت‌های اقتصادی تولیدی به‌منظور ایجاد فضای رقابتی عادلانه‌تر.<br>• اقدام در زمینهٔ کاهش ریسک‌های اقتصادی و اعتباری و امن‌تر نمودن فضای کسب و کار. | • افزایش سهم نیروی انسانی متخصص در فرآیند تولید و ارتقاء سطح سرمایه انسانی و استفادهٔ بهینه از آن.[3]<br>• ایجاد مکانیزم‌هایی در راستای افزایش انگیزهٔ نیروی کار در جهت تشویق نوآوری، خلاقیت، افزایش دقت و جدّیت افراد.[4]<br>• ارتقاء دانش فنی از طریق افزایش سهم هزینه‌های تحقیق و توسعه، ارتقاء کارایی آن و جذب اثرات سرریز تحقیق و توسعه خارجی.<br>• استفاده بیشتر از فناوری اطلاعات و ارتباطات(*ICT*) در فرآیندهای اکتشاف، آماده‌سازی، استخراج و کانه‌آرایی.[5]<br>• استفادهٔ بهینه از انرژی و نهاده‌های واسطه‌ای و بهبود در کیفیت نهاده‌های واسطه‌ای.<br>• کاهش سن متوسط تجهیزات و امکانات سرمایه‌ای.<br>• گسترش میزان ارائهٔ اطلاعات پایه معدنی و تکمیل بانک اطلاعات زمین‌شناسی و اکتشافی کشور.<br>• گسترش فرآیندها و فعالیت‌هایی که ماهیتاً ارزش‌افزوده بیشتری را برای بخش معدن ایجاد نموده و از خام‌فروشی مواد معدنی جلوگیری می‌نمایند، نظیر فرآیند کانه‌آرایی.<br>• تمرکز طراحی روش‌های استخراج معادن روی روش‌های استخراج سطحی[6]، و یا تمرکز روی روش‌های استخراج زیرزمینی که بهره‌وری بیشتری دارند.[7] |

## منابع فارسی

[1]. فرآیند تولید در بخش معدن ماهیتاً سرمایه‌بر بوده و تنها از طریق بکارگیری تکنولوژی‌های سرمایه‌بر می‌توان به سطوح بالای بهره‌وری دست یافت.

[2]. خصوصی‌سازی به‌معنای واگذاری مدیریت به بخش خصوصی نه به‌معنای صرفاً واگذاری مالکیت به بخش خصوصی. به عبارت دیگر باید سهام مدیریتی بنگاه‌های اقتصادی به دست بخش خصوصی سپرده شود.

[3]. تحصیلات، تنها در صورتی به رشد ارزش‌افزوده بیشتر می‌انجامد که از افراد تحصیلکرده در جایی که موجب افزایش بهره‌وری در تولید می‌گردد، استفاده شود. بنابراین، تلاش در جهت کاربردی نمودن تحصیلات و بکارگیری دانش‌آموختگان در تخصص‌های تحصیلی خود، از شرط‌های تأثیرگذاری مثبت سرمایه انسانی بر بهره‌وری کل عوامل و رشد ارزش‌افزوده است.

[4]. به‌عنوان مثال، تدوین یک نظام تعیین دستمزد مبتنی بر بهره‌وری نیروی کار به منظور افزایش انگیزه نیروی کار برای کار بیشتر و مفیدتر، ابداع و نوآوری، خلاقیت و استفاده بهینه از امکانات سرمایه‌ای.

[5]. اسمیت(۲۰۰٤) نشان می‌دهد که استفاده از فناوری اطلاعات و ارتباطات در معادن زغال‌سنگ کانادا باعث ارتقاء سطح بهره‌وری در معادن مذکور شده است.

[6]. هر چند که انتخاب روش استخراج مناسب، تابع پارامترهای فنی و اقتصادی متنوعی می‌باشد، ولیکن تحقیقات اقتصادی نظیر دارمستادتر(۱۹۹۷) و اسمیت(۲۰۰٤) بیانگر این موضوع هستند که تولید در معادن سطحی بهره‌وری بالاتری از تولید در معادن زیرزمینی به همراه دارد. لذا تلاش در راستای طراحی استخراج به صورت سطحی(هرچند که احتمالاً نسبت باطله به مادهٔ معدنی را افزایش می‌دهد) ممکن است از نظر اقتصادی و به‌ویژه شاخص بهره‌وری، تلاشی پرفایده باشد.

[7]. به‌عنوان مثال، دارمستادتر(۱۹۹۷) نشان داده است که بهره‌وری در روش‌های استخراج پیوسته و به‌ویژه روش استخراج جبهه‌کار بلند(*Long Wall*) در معادن زغال‌سنگ ایالات متحده، بهره‌وری بیشتری از سایر روش‌های استخراج داشته است.

## منابع لاتین